\documentclass{article}

\usepackage{microtype}
\usepackage{graphicx}
\usepackage{subfigure}
\usepackage{booktabs} 

\usepackage{hyperref}

\usepackage{listings}

\usepackage[accepted]{icml2025}

\usepackage{amsmath}
\usepackage{amssymb}
\usepackage{mathtools}
\usepackage{amsthm}

\usepackage[capitalize,noabbrev]{cleveref}

\usepackage{threeparttable}
\usepackage{multirow}
\usepackage{booktabs}
\usepackage{xcolor}
\usepackage{bbding}

\usepackage{booktabs}       
\usepackage{amsfonts}       
\usepackage{nicefrac}       
\usepackage{microtype}      
\usepackage{xcolor}         
\usepackage{graphicx}
\usepackage{multirow}
\usepackage{colortbl}
\usepackage{booktabs}
\usepackage{xspace}
\usepackage{enumitem}
\usepackage{array}
\usepackage{pifont}  
\usepackage{wrapfig}

\usepackage{xcolor}

\theoremstyle{plain}

\theoremstyle{definition}

\theoremstyle{remark}

\usepackage[textsize=tiny]{todonotes}

\newcommand{\modelname}{RefineNovo}

\icmltitlerunning{Curriculum Learning for Biological Sequence Prediction: The Case of De Novo Peptide Sequencing}

\begin{document}

\twocolumn[
\icmltitle{Curriculum Learning for Biological Sequence Prediction:\\ The Case of De Novo Peptide Sequencing}



\icmlsetsymbol{equal}{*}
\icmlsetsymbol{intern}{$\dagger$}

\begin{icmlauthorlist}
\icmlauthor{Xiang Zhang}{equal,intern,fff,xxx}
\icmlauthor{Jiaqi Wei}{equal,yyy,zzz}
\icmlauthor{Zijie Qiu}{equal,yyy,fff}
\icmlauthor{Sheng Xu}{yyy,fff}

\icmlauthor{Nanqing Dong}{yyy}
\icmlauthor{Zhiqiang Gao}{yyy}
\icmlauthor{Siqi Sun}{fff,yyy}
\end{icmlauthorlist}

\icmlaffiliation{fff}{Fudan University}
\icmlaffiliation{xxx}{University of British Columbia}
\icmlaffiliation{yyy}{Shanghai Artificial Intelligence Laboratory}
\icmlaffiliation{zzz}{Zhejiang University}

\icmlcorrespondingauthor{Xiang Zhang}{xzhang23@ualberta.ca}
\icmlcorrespondingauthor{Siqi Sun}{siqisun@fudan.edu.cn}

\icmlkeywords{Machine Learning, ICML}
\vskip 0.3in]



\printAffiliationsAndNotice{\icmlEqualContribution$^\dagger$work done while interning at Fudan University.} 

\begin{abstract}
Peptide sequencing—the process of identifying amino acid sequences from mass spectrometry data—is a fundamental task in proteomics. Non-Autoregressive Transformers (NATs) have proven highly effective for this task, outperforming traditional methods. Unlike autoregressive models, which generate tokens sequentially, NATs predict all positions simultaneously, leveraging bidirectional context through unmasked self-attention.
However, existing NAT approaches often rely on Connectionist Temporal Classification (CTC) loss, which presents \textbf{\textit{significant}} optimization challenges due to CTC's complexity and increases the risk of training failures. To address these issues, we propose an improved non-autoregressive peptide sequencing model that incorporates a structured protein sequence \textbf{\textit{curriculum learning}} strategy. This approach adjusts protein's learning difficulty based on the model’s estimated protein generational capabilities through a sampling process, progressively learning peptide generation \textit{from simple to complex sequences}. 
Additionally, we introduce a self-refining inference-time module that iteratively enhances predictions using learned NAT token embeddings, improving sequence accuracy at a fine-grained level. \textbf{Our curriculum learning strategy reduces NAT training failures frequency by more than 90\%} based on sampled training over various data distributions. Evaluations on nine benchmark species demonstrate that our approach \textbf{outperforms all previous methods across multiple metrics and species}. Model and source code are available at \href{https://github.com/BEAM-Labs/denovo}{Github}.
\end{abstract}

\section{Introduction }

\begin{figure}[tb]
\centering
\includegraphics[width=0.9\linewidth, height=0.22\textheight]{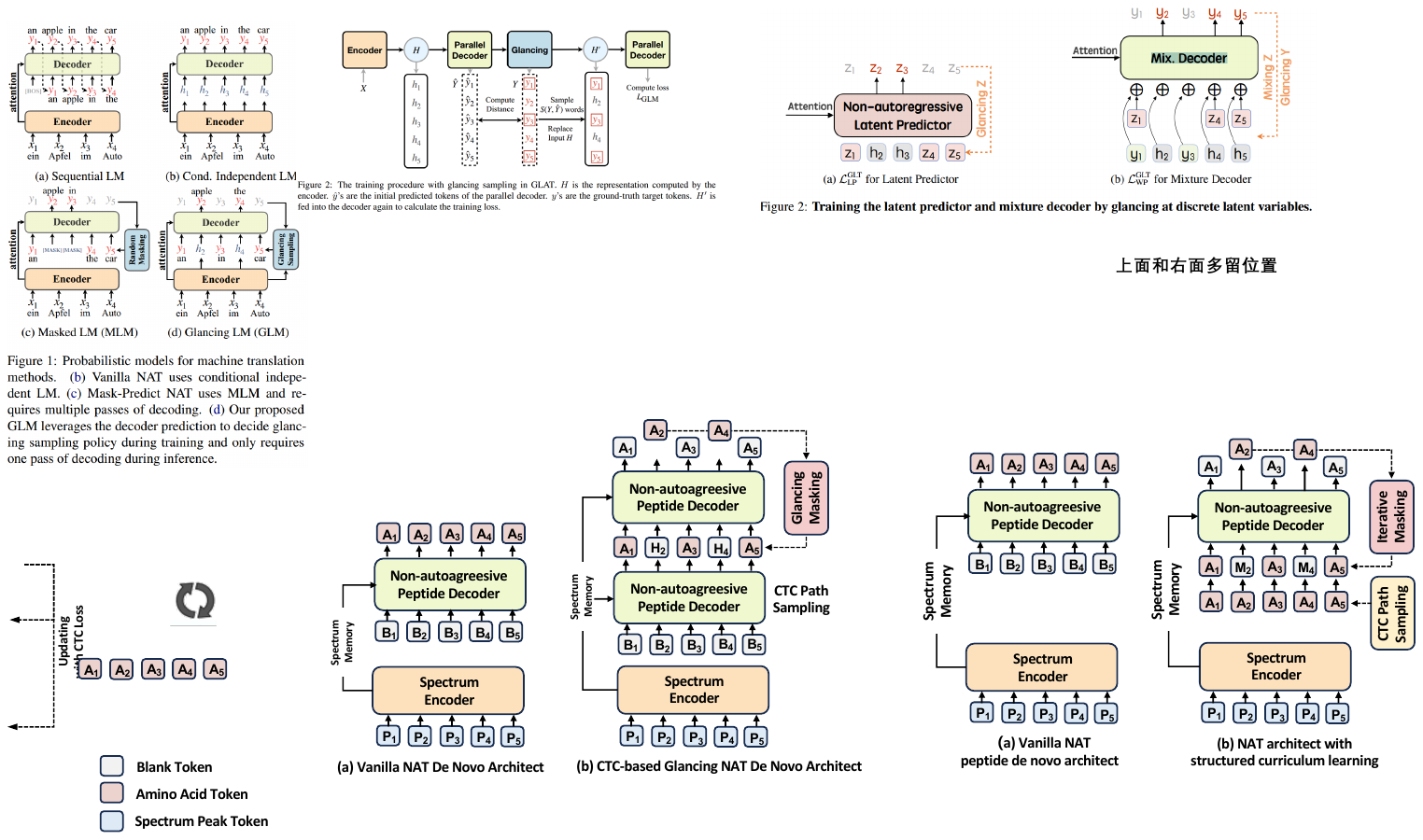}
\vspace{-1em}
\caption{Comparison between (a) Vanilla NAT and (b) our proposed NAT peptide sequencing architectures, which integrate curriculum learning and a self-refining module.}

\label{fig:small}
\end{figure}

\begin{figure}[tb]
\centering
\includegraphics[width=1\linewidth, height=0.16\textheight]{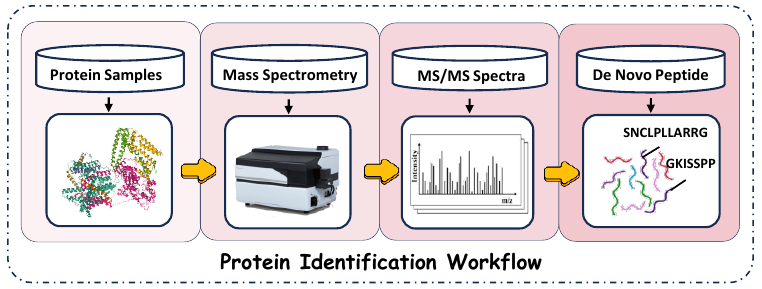}
\vspace{-1em}
\caption{Overview of the protein identification workflow. Protein samples are digested into peptides, which are then analyzed using mass spectrometry to generate MS/MS spectra. These spectra are subsequently used in de novo peptide sequencing to determine the peptide sequences.}
\label{fig:process}
\end{figure}

Peptide sequencing via tandem mass spectrometry plays a pivotal role in proteomics research, with significant implications for fundamental and applied studies in chemistry, biology, medicine, and pharmacology~\cite{aebersold2003mass,ng2023algorithms}. The principal aim of peptide sequencing, as illustrated in Figure \ref{fig:process}, is to deduce the amino acid sequences of segmented short protein sequences from mass spectra derived from specific biological samples. Database-searching-based sequencing and de novo peptide sequencing are the two most widely used methods for identifying peptide sequences~\cite{chen2020bioinformatics}. Traditional database search methods have inherent limitations, such as the inability to identify peptide sequences absent from the database and the dramatic increase in computational costs and processing times as the database expands. In contrast, de (from) novo (scratch)  peptide sequencing directly deduces peptide sequences from the spectra, overcoming the limitations of the static search space in database-based algorithms~\cite{muth2018potential}.

Deep learning-based models~\cite{lecun2015deep,liu2022character,zhang2023pi} have revolutionized the scientific field, including de novo sequencing. Notably, non-autoregressive Transformer~\cite{gu2017non,xiao2023survey} models have demonstrated superior performance among all deep learning-based methods in protein sequence predictions~\cite{lin2023evolutionary,hayes2024simulating,zhang2024pi}. Unlike autoregressive models, which rely on ``next token prediction'' for generation, non-autoregressive models compute the token probabilities at each position simultaneously. This parallel prediction approach enables bidirectional information flow during generation through a self-attention mechanism. Specifically, each position accesses information from all surrounding positions when generating its token, as opposed to relying solely on the preceding ones in an autoregressive model.
Such an approach closely aligns with the nature of protein formation and has proven to be more accurate and efficient in de novo sequencing as well as some of the other bio-sequence-related tasks~\cite{eloff2023novo,zhang2024pi}.

Given that straightforward token-by-token optimization with cross-entropy loss in Non-Autoregressive (NAT) modeling often leads to poor global sequence-level connectivity~\cite{gu2017non} (explained in later section), many NAT-based models employ alternative loss objectives such as Connectionist Temporal Classification (CTC) loss \cite{graves2006connectionist,graves2014towards} and Directed Acyclic Transformer (DAT) loss \cite{huang2022directed}. The previous NAT-based de novo sequencing model (Figure \ref{fig:small}a) also utilized the CTC objective and has achieved state-of-the-art performance (by 10\% improvement compared to AT models). However, the naive CTC objective is far from flawless. It involves complex reduction rules (detailed in later section) and creates a vast search space during optimization, leading to an \textit{unstable learning curve} and \textit{slow convergence} during training (Figure \ref{fig:case}). This optimization challenge adversely affects overall performance.

In this work, we enhance standard CTC training and introduce the first protein curriculum learning model for improved sequence prediction. Our method exposes the NAT model to proteins of varying difficulty levels, guiding it to progressively learn from simple to complex sequences, mirroring the human learning process by starting with easier prediction targets and gradually increasing difficulty.

Specifically, we integrate \textit{curriculum learning} with a CTC-based training objective, sampling from the model’s own CTC output path during training to determine learning targets and adjust difficulty based on its \textit{current performance} (Figure \ref{fig:small}b). By exposing the model to simpler targets first, the curriculum learning approach enables a more gradual and smooth learning process, eliminating the need for the model to grapple with complex protein generation rules from the very beginning.

Furthermore, iterative refinement has proven highly effective in various protein modeling tasks, including protein language modeling and structure prediction \cite{jumper2021highly, abramson2024accurate}. By incorporating previously predicted sequences as input, models can iteratively refine their predictions, improving generation quality. However, traditional NAT models \cite{zhang2024pi} struggle with this approach, as they are trained without conditioning on input sequences beyond positional encoding.
Our methodology overcomes this limitation by incorporating input sequence information into the curriculum learning process. By leveraging learned token embeddings, we seamlessly integrate iterative refinement during inference, significantly enhancing the accuracy of protein sequence generation.

Experiments and case studies demonstrate that our training framework effectively mitigates three common failure modes in NAT protein models: loss explosion, extreme over-fitting, and unstable loss convergence. Our approach ensures a consistently smooth training curve across diverse datasets, including both well-distributed and poorly distributed samples, achieving over 90\% reduction in training failures.

Experiments conducted on the widely recognized 9-species-V1 \cite{tran2017novo} and 9-species-V2 \cite{yilmaz2023sequence} benchmark datasets demonstrate the superiority of the proposed {\modelname} over all previous models across various evaluation metrics, establishing a new state-of-the-art in the field. 
Additionally, we investigate {\modelname}'s capacity to differentiate between amino acids with similar masses, showcasing its adeptness in discerning subtle distinctions among challenging amino acids.
The overall performance again highlights the exceptional capabilities of {\modelname}, positioning it as a promising and innovative tool in the field of proteomics. 

\section{Related Work}

\noindent \textbf{Autoregressive v.s. Non-Autoregressive.} 
Autoregressive generation with Transformer models, where tokens are predicted sequentially, has demonstrated outstanding performance across various scenarios~\cite{xiao2023survey}. However, this approach can be time-consuming, especially for long sequences. To address this limitation, \citet{gu2017non} introduced the first NAT model for neural machine translation, which facilitates parallel decoding and significantly accelerates the generation process. This substantial increase in inference speed has attracted much attention to NAT methods, leading to impressive progress through techniques such as knowledge distillation~\cite{zhou2020improving,ding2021rejuvenating,shao2022one}, innovative learning strategies~\cite{qian2020glancing, ding2021progressive,zhu2022non}, iterative methods~\cite{stern2019insertion,guo2020jointly,savinov2021step,huang2022improving}, latent variable-based techniques~\cite{ma2019flowseq,shu2020latent,song2021alignart,bao2021non}, enhancement techniques~\cite{wang2019non,yin2023ttida,guo2019non,ding2020context,liu2022character,huang2022non}, and different criterion~\cite{saharia2020non,shao2020minimizing,ghazvininejad2020aligned}, all of which have further improved the performance of NAT models.

Building on these advancements, NAT models have also shown substantial benefits in biology-related sequence generation tasks, such as protein generative modeling~\cite{lin2023evolutionary,hayes2024simulating} and de novo peptide sequencing~\cite{zhang2024pi}, demonstrating the powerful generative ability brought by the bi-directional scheme.

\noindent \textbf{Deep Learning-Based De Novo Peptide Sequencing.}
With the advent of deep learning~\cite{lecun2015deep,gao2023deep,jin2023contranovo}, the performance of de novo peptide sequencing has markedly improved. A notable pioneer in this domain is DeepNovo~\cite{tran2017novo}, the first deep learning-based method leveraging CNNs and LSTMs to model spectra and peptide sequences. Following this innovation, numerous deep learning-based approaches have emerged for de novo peptide sequencing~\cite{zhou2017pdeep,Karunratanakul2019,yang2019pnovo,liu2023accurate,mao2023mitigating}. However, these methods often involve complex modeling techniques, including the combination of multiple neural networks and intricate post-processing steps~\cite{yilmaz2022novo}.
Casanovo \citep{yilmaz2022novo,yilmaz2023sequence} and its derivatives (e.g., AdaNovo \citep{xia2024adanovo}, HelixNovo \citep{yang2024introducing}, InstaNovo \citep{eloff2023novo}, SearchNovo~\citep{xia2024bridging}), and RankNovo~\cite{qiu2025universal} introduced the Transformer architecture to directly model the de novo peptide sequencing problem analogously to machine translation in natural language processing. Among these, ContraNovo~\cite{jin2024contranovo} implemented a contrastive learning strategy and integrated additional amino acid mass information, thereby enhancing sequencing accuracy. Recently, PrimeNovo~\cite{zhang2024pi} presented the first non-autoregressive Transformer architecture, achieving state-of-the-art results in this task.

PrimeNovo highlights the potential of non-autoregressive decoding in de novo peptide sequencing. In this study, we aim to build on this potential by refining the NAT-based method to further enhance de novo sequencing performance. Our focus is on increasing the reliability of the decoding process, which we believe will lead to advancements in the field.

\begin{figure*}[tb]
\centering
\includegraphics[width=0.93\textwidth, height=0.395\textheight]{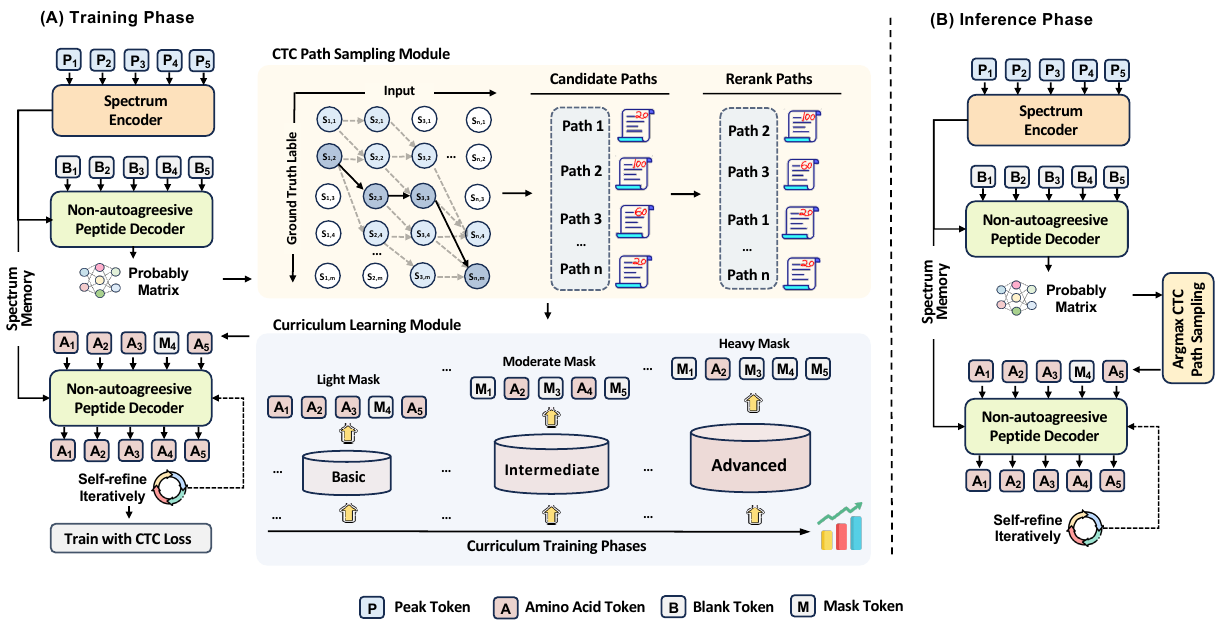}
\caption{The architecture of {\modelname}. (A) The training phase of {\modelname} begins with a Spectrum Encoder that processes the input spectra. Following encoding, the Non-Autoregressive Peptide Decoder leverages the encoded spectra and blank tokens to generate a probability matrix. To predict potential sequences, CTC path sampling identifies all candidate paths, which are then re-ranked for optimal selection. The model employs a  Masking strategy, progressively masking tokens to support a curriculum training approach that advances from basic to advanced stages. Further refining the learning process, the model iteratively adjusts based on previously predicted sequences and is trained using the CTC loss function.
(B) In the inference phase, the model operates similarly to the training phase, with the primary difference being the use of Argmax CTC path sampling. To maintain accuracy, the model continues to iteratively refine its predictions, ensuring precise peptide sequence generation.} 
\label{fig:modelArchi}
\end{figure*}

\section{Method}

\subsection{Problem Formulation}
We first formally define the task of de novo peptide sequencing. Our goal is to translate a given mass spectrometry spectrum into the sequence of amino acids it encodes (Figure \ref{fig:process}). The spectrum consists of a set of paired mass-to-charge ratio values (x-axis) with corresponding peak intensity values (y-axis), denoted as $\mathcal{I}= \{(\text{mz}^{(1)}, \text{p}^{(1)}), (\text{mz}^{(2)}, \text{p}^{(2)}), \ldots, (\text{mz}^{(k)}, \text{p}^{(k)})\}$. Additionally, two more pieces of information about the target peptide are provided by the mass spectrometer: the overall peptide mass (precursor mass) $m$ and the entire peptide charge $z$. Using all these inputs, we aim to predict the correct amino acid sequence $\mathcal{A} = (a_1, a_2, \ldots, a_n)$.
\subsection{Non-autoregressive Transformer BackBone }
Our model utilizes an encoder-decoder transformer architecture, similar to the previous NAT-based de novo model \cite{zhang2024pi}. To incorporate the curriculum learning strategy, we make several architectural changes, outlined in detail below.

\textbf{Spectrum Encoder.} The encoder compresses the spectrum input $\mathcal{I}$ into a meaningful embedding $\mathbf{E}$ in latent space. We treat the values in $\mathcal{I}$ as a sequence and encode each float value of $\text{mz}^{(i)}$ using a sinusoidal encoding as follows:: 
\begin{equation}
\resizebox{0.9\hsize}{!}{$\text{e}^{0}_{i} (\text{mz}) = \begin{cases}\sin((\text{mz})/(\frac{(\text{mz})_{\max}}{(\text{mz})_{\min}}(\frac{(\text{mz})_{\min}}{2\pi})^{\frac{2i}{d}})),&\text{for}~i\leq \frac{d}{2} \\\cos((\text{mz})/(\frac{(\text{mz})_{\max}}{(\text{mz})_{\min}}(\frac{(\text{mz})_{\min}}{2\pi})^{\frac{2i}{d}})),&\text{otherwise} \end{cases}$}
\end{equation}
where $d$ is the hidden dimension of our model. Similarly, the peak intensity $\text{p}^{(i)}$ is encoded with the same function into $d$ dimensions and then added to $\text{e}(\text{mz})$ at each position.  The encoded spectrum embedding $\mathbf{E}^{0} = (\text{e}^0_1, \text{e}^0_2, \cdots, \text{e}^0_k)$ is then fed into the Transformer Encoder with $m$ layers, where the $j$th encoder layer updates the last layer embedding $\mathbf{E}^{j-1}$ as follows:
\begin{equation}
    \mathbf{E}^{j} = \text{SelfAttention}(\text{e}^{j-1}_0, \text{e}^{j-1}_1, \cdots, \text{e}^{j-1}_k )
\end{equation}
The output from the last layer, $\mathbf{E}^{(m)}$, is used as the feature representation of the input spectrum and will be used in the decoding process.

\textbf{Peptide Decoder.} Unlike autoregressive decoders, which use self-attention for next token prediction, NAT decoders generate token probabilities for each position independently, using a self-attention-based decoder module. Unlike previous NAT designs that only take positional encodings as input, our model incorporates calculated sequences as input for curriculum learning purpose. To accommodate this, we add an embedding layer, denoted as $\mathbf{h}^{0}_i = \text{EmbeddingLayer}(\text{y}_i)$, where $\text{y}_i$ is the input token at the $i$th position. The encoded input is then passed through both self-attention and cross-attention layers with the spectrum features $\mathbf{E}^{m}$. Finally, the output of the last layer, $\mathbf{h}^{(L)}$, is mapped to the probability of tokens as: $P_s(\cdot \mid \mathcal{I}) = \text{softmax}(W \mathbf{h}_s^{(L)})$ for decoding position $s$.

\textbf{CTC Training.}  Using a naive cross-entropy loss in parallel prediction models can lead to the multi-modal problem\footnote{For example, translation of "au revoir" to english might result in sentence such as "good you" or "see bye". } due to the lack of token dependencies~\cite{gu2017non}. To address this, we use the Connectionist Temporal Classification (CTC) loss as our optimization objective. In CTC, we first define a maximum generational length $T$, which can be reduced to the target length using the following rules $\Gamma(\cdot)$: 1) consecutive identical tokens are merged; 2) placeholder token $\epsilon$ is removed; 3) identical tokens adjacent to $\epsilon$ are not merged. For example, the sequence AABC$\epsilon$C is reduced to ABCC.

During training, instead of maximizing the generation probability of the true token $a_i$ at position $i$, CTC maximizes the probability of all generation paths $\mathbf{y} = (\text{y}_1, \text{y}_2, \ldots, \text{y}_T)$ such that $\Gamma(\mathbf{y}) = \mathcal{A}$. For example, with the target sequence ATC and a generational length of 5, decoding paths such as AATTC and AA$\epsilon$TC will be assigned higher probabilities as they can be reduced to the true sequence. Conversely, the path A$\epsilon$ATTC will be discouraged since it does not reduce to the target sequence.

Therefore, the objective for CTC is to maximize the total probability $P(A|\mathcal{I})$ of all valid paths (those that map to the target sequence $A$), or equivalently, to minimize the negative logarithm of this probability:
\begin{align}
  \mathcal{L}_{CTC} &= - \log P(A | \mathcal{I}) \notag 
  = - \log \left( \sum_{\mathbf{y}:\Gamma(\mathbf{y})=A} P(\mathbf{y}|\mathcal{I}) \right) \label{eq:ctc_loss_corrected}
\end{align}
The term $P(\mathbf{y}|\mathcal{I})$ represents the probability of a single alignment path $\mathbf{y}$. If this path probability is computed as a product of individual token probabilities (e.g., $P(\mathbf{y}|\mathcal{I}) = \prod_{\mathrm{y}_i \in \mathbf{y}} P(\mathrm{y}_i|\mathcal{I})$ for tokens $\mathrm{y}_i$ in path $\mathbf{y}$), then its logarithm, $\log P(\mathbf{y}|\mathcal{I})$, is indeed equal to the sum of the individual log token probabilities, $\sum_{\mathrm{y}_i \in \mathbf{y}} \log P(\mathrm{y}_i|\mathcal{I})$. This is the property that 'the log of the product of all token probabilities equals the sum of the log of each token’s probability.' It's important to note that this property applies to the calculation of the log-probability of a \textit{single path}, whereas the CTC loss involves summing the actual probabilities $P(\mathbf{y}|\mathcal{I})$ of \textit{all} valid paths \textit{before} taking the logarithm.
The detailed explanation of CTC loss calculation is in Appendix Section \ref{appendix:ctc_loss}.

\textbf{PMC Unit.} Following the work of Zhang et al.~\cite{zhang2024pi}, we apply a precise mass control post-decoding unit. Since the precursor mass $m$ provided as input is a strict constraint for the generated sequence $\mathbf{y}$, we use a dynamic programming solver to find the path that satisfies this constraint. Specifically, this is modeled as a knapsack problem where the total generation mass $m$ is the maximum capacity of the knapsack. We select items (amino acid tokens) from each position to fill the bag, with each token having a value (predicted log probability) and a weight (molecular mass). The most valuable (probable) path that meets the mass constraint is determined through a 2D dynamic programming table.
The detailed decoding algorithm can be referred to in Appendix Section \ref{appendix:pmc}.

\subsection{Protein Curriculum Learning with CTC-based NAT} Parallel generative models face more complex optimization landscapes than autoregressive models due to their larger search space. In autoregressive models, predictions at time step 
t
t are conditioned on previous tokens, i.e.,$P$($x_t | x_{1:t-1}, \mathcal{I}$), effectively constraining the search space based on prior information $x_{1:t-1}$ and simplifying optimization.  In contrast, NAT models predict $P$($x_t$) independently, without conditioning on previous outputs, leading to a significantly larger search space that makes convergence to the global optimum more challenging. Additionally, the implicit reduction rule in CTC further complicates learning, making target path discovery more difficult than next-token prediction in autoregressive models. As a result, we frequently observe instability in loss convergence and high training failure rates in NAT models (Figure \ref{fig:case}).

To address these challenges, we introduce a curriculum learning strategy tailored for CTC-based protein prediction.  
Unlike naive NAT models, which learn to predict the entire sequence from scratch, our model selectively masks parts of the target sequence$\mathcal{A}$ with a special masking token. This modifies the learning objective from independent sequence prediction to conditioned probability estimation:
\begin{equation}
    \mathcal{L} = P(\mathcal{A} | \rho (\mathcal{A}, \mathbf{y}), \mathcal{I})
\end{equation} 
where $\rho(\mathcal{A}, \mathbf{y})$ represents the selected unmasked tokens. By "leaking" some ground-truth tokens, prediction becomes easier, and the search space is effectively reduced as:
\begin{equation}
    \text{SPACE}(NAT) \sim \gamma \cdot \text{TYPE}(Loss) \cdot \rho_\text{ratio} 
\end{equation}
In this case, SPACE$(NAT)$ reaches zero when no masking is applied (i.e. $\rho_\text{ratio}=0$ , the true label is fully provided) and is maximized when all positions are masked  (i.e. Naive NAT model, $\rho_\text{ratio}=1$). The TYPE(ctc) yield much larger search space than TYPE(cross entropy) since each target sequence requires searching through $O(nT)$ pre-ctc-reduction paths, rather than a single path as in cross-entropy loss.

However, while this strategy is straightforward for non-CTC-based learning objectives, it becomes significantly more complex when CTC reduction rules are involved. In CTC, position $t$ in the generated sequence does not necessarily correspond to $a_t$ in the true label as the CTC path is often much longer before reduction. This makes it impractical to create our conditional input $\rho (\mathcal{A}, \mathbf{y})$. To address this, we propose a CTC-based curriculum strategy.

First, we perform a forward pass with an \textbf{empty} decoder input to obtain the probability distribution for each position $t \in (1, \cdots, T)$. We then calculate the sequence probability for all valid CTC decoding paths $\mathbf{y}$ such that $\Gamma(\mathbf{y}) = \mathcal{A}$. We re-rank all such paths $\mathbf{y}$ according to their total sequence probability and choose the most likely path $\mathbf{y}'$ as our to-be-masked input $\mathbf{y}'$ (Figure \ref{fig:modelArchi} Yellow). We apply masking to obtain the input sequence $\rho(\mathcal{A}, \mathbf{y}')$  which is fed into the decoder for conditioned generation and backpropagation update.  

This approach exposes the model to partial information from one valid decoding path, allowing it to infer the structure of the true CTC path while learning to predict masked tokens and generalize to unobserved paths. Intuitively, this partial sequence exposure lowers learning difficulty by strategically “leaking” information, guiding the model to internalize the alignment patterns inherent in CTC decoding.

\textbf{Difficulty Annealing.} Naively applying masking (e.g. $\rho_\text{ratio} = 0.5$) to reduce learning difficulty can result in ineffective learning due to excessive information leakage. To address this, we adopt a difficulty annealing approach to adaptively reduce the masking ratio over the course of training based on model's per-epoch performance. Specifically, during the forward pass in training time, we calculate the model's prediction accuracy $\text{acc}(\mathcal{A}, \mathbf{y}^{\text{argmax}})$ based on CTC-argmax decoding. The masking ratio is then calculated as $\rho_\text{ratio} = \alpha (1 - \text{acc}(\mathcal{A}, \mathbf{y}^{\text{argmax}}))$. Consequently, \textbf{most tokens will be unmasked at the beginning of training, with more tokens being masked as the model's predictions become more accurate over time}.

We present the curriculum learning method in Algorithm~\ref{alg:ctc_curriculum}. This algorithm outlines the key steps for computing alignment paths, selecting oracle tokens, and applying adaptive masking for training.

\begin{algorithm}[t]
    \caption{Curriculum Learning with CTC-based NAT}
    \label{alg:ctc_curriculum}
    \begin{algorithmic}[1]
        \REQUIRE Training batch $\mathbf{batch}$, decoder $\mathbf{decoder}$, peek factor $\alpha$
        \ENSURE Masked oracle sequence $\mathbf{glat\_prev}$

        \IF {Training mode}
            \STATE \textcolor{blue}{\# Forward pass without gradient updates}
            \STATE $\mathbf{word\_ins\_out}, \mathbf{tgt\_tokens} \leftarrow \text{forward\_step}(\mathbf{batch})$
            \STATE $\mathbf{pred\_tokens} \leftarrow \arg\max(\mathbf{word\_ins\_out})$
            
            \STATE \textcolor{blue}{\# Compute best alignment paths via CTC}
            \STATE $\mathbf{best\_aligns} \leftarrow \text{CTC\_align}(\mathbf{word\_ins\_out}, \mathbf{tgt\_tokens})$
            \STATE $\mathbf{oracle\_pos} \leftarrow \text{midpoint}(\mathbf{best\_aligns})$
            \STATE $\mathbf{oracle} \leftarrow \text{gather}(\mathbf{tgt\_tokens}, \mathbf{oracle\_pos})$
            \STATE $\mathbf{oracle\_masked} \leftarrow \text{mask}(\mathbf{oracle}, \text{blank where needed})$

            \STATE \textcolor{blue}{\# Compute model accuracy via CTC-argmax decoding}
            \STATE $\mathbf{same\_num} \leftarrow \sum (\mathbf{pred\_tokens} == \mathbf{oracle\_masked})$
            \STATE $\mathbf{seq\_lens} \leftarrow |\mathbf{pred\_tokens}|$
            \STATE $\mathbf{acc} \leftarrow 1 - \frac{\mathbf{same\_num}}{\mathbf{seq\_lens}}$

            \STATE \textcolor{blue}{\# Compute adaptive masking probability based on performance}
            \STATE $\rho_\text{ratio} \leftarrow \alpha (1 - \mathbf{acc})$

            \STATE \textcolor{blue}{\# Apply curriculum masking}
            \STATE $\mathbf{mask} \leftarrow \text{random}() < \rho_\text{ratio}$
            \STATE $\mathbf{glat\_prev} \leftarrow \text{mask}(\mathbf{oracle\_masked}, \mathbf{mask})$

        \ENDIF
        
        Return $\mathbf{glat\_prev}$
    \end{algorithmic}
\end{algorithm}

The python implementation as well as pseudo-code of this learning can be seen in Appendix Sec. \ref{app:curriculum} and Algorithm \ref{alg:ctc_curriculum}.

\subsection{CTC-Curriculum-based Protein Iterative Refinement}
Our trained model includes a curriculum token embedding layer designed to encode conditional inputs, $\rho(\mathcal{A}, \mathbf{y}')$. However, during inference, this layer is largely unused since there is no ground-truth label, and all conditional inputs are mask-tokens. To maximize its utility, we adopt a multi-pass forward approach to iteratively refine the generated sequence. Notably, prior studies have demonstrated that iterative refinement enhances prediction accuracy in protein-related tasks.

To this end, we propose a whole-sequence refinement inference scheme using the previously trained CTC embedding layer, which can encode meaningful information from any CTC path $\mathbf{y}$. Specifically, at the $i$-th iteration, the previously decoded pseudo path using argmax, $\mathbf{y}^{(i-1)}$, is used as input for the decoder again for the conditional generation:

\begin{equation}
\mathbf{y}^{(i)} = \text{argmax} \ \ P(\cdot | \mathcal{I}, \text{EmbeddingLayer}(\mathbf{y}^{(i-1)}))
\end{equation}

This process is iterated for $N$ times, with the final output $P(\cdot| \mathcal{I}, \mathbf{y}^{(N-1)})$ sent to the PMC unit for final decoding. We perform whole-sequence recycling instead of decoding partial tokens in each pass because the PMC unit requires a probability distribution for all positions from a single pass to avoid distributional shifts. The algorithm detail is in Appendix Section \ref{app:iter}.

\begin{table*}[t]
\setlength{\belowcaptionskip}{2mm}
\centering
\begin{threeparttable}
\setlength{\tabcolsep}{0.8mm} 
\renewcommand{\arraystretch}{1.2} 
\scalebox{0.8}{
\begin{tabular}{c|c|l|ccccccccc|c}
\toprule
Metrics & Category & Methods & Mouse & Human & Yeast & M. mazei & Honeybee & Tomato & Rice bean & Bacillus & C. bacteria & \textbf{Average} \\
\midrule
\multirow{8}{*}{AA Precision} & DB & Peaks Novo~\cite{ma2003peaks} & 0.600 & 0.639 & 0.748 & 0.673 & 0.633 & 0.728 & 0.644 & 0.719 & 0.586 & 0.663 \\
\cline{2-13}
& \multirow{6}{*}{AR} & Deep.~\cite{tran2017novo} & 0.623 & 0.610 & 0.750 & 0.694 & 0.630 & 0.731 & 0.679 & 0.742 & 0.602 & 0.673 \\
&& Point.~\cite{qiao2021computationally} & 0.626 & 0.606 & 0.779 & 0.712 & 0.644 & 0.733 & 0.730 & 0.768 & 0.589 & 0.687 \\
&& Casa.~\cite{yilmaz2022novo} & 0.689 & 0.586 & 0.684 & 0.679 & 0.629 & 0.721 & 0.668 & 0.749 & 0.603 & 0.667 \\
&& Ada.~\cite{xia2024adanovoadaptiveemphdenovo} & 0.646 & 0.618 & 0.793 & 0.728 & 0.650 & 0.740 & 0.719 & 0.739 & 0.642 & 0.697 \\
&& Casa.V2~\cite{yilmaz2023sequence} & 0.760 & 0.676 & 0.752 & 0.755 & 0.706 & 0.785 & 0.748 & 0.790 & 0.681 & 0.739 \\

\cline{2-13}
& \multirow{2}{*}{NAR} & Prime.~\cite{zhang2024pi} & 0.784 & 0.729 & {0.802} & {0.801} & {0.763} & {0.815} & {0.822} & {0.846} & {0.734} & {0.788} \\ 
&& \cellcolor{blue!10}\textbf{Ours} & \cellcolor{blue!10}\textbf{0.800} & \cellcolor{blue!10}\textbf{0.730} & \cellcolor{blue!10}\textbf{0.818} & \cellcolor{blue!10}\textbf{0.819} & \cellcolor{blue!10}\textbf{0.780} & \cellcolor{blue!10}\textbf{0.825} & \cellcolor{blue!10}\textbf{0.835} & \cellcolor{blue!10}\textbf{0.854} & \cellcolor{blue!10}\textbf{0.742} & \cellcolor{blue!10}\textbf{0.800} \\ 
\midrule
\multirow{8}{*}{Peptide Recall} & DB & Peaks Novo~\cite{ma2003peaks} & 0.197 & 0.277 & 0.428 & 0.356 & 0.287 & 0.403 & 0.362 & 0.387 & 0.203 & 0.322 \\ 
\cline{2-13}
& \multirow{6}{*}{AR} & Deep.~\cite{tran2017novo} & 0.286 & 0.293 & 0.462 & 0.422 & 0.330 & 0.454 & 0.436 & 0.449 & 0.253 & 0.376 \\
&& Point.~\cite{qiao2021computationally} & 0.355 & 0.351 & 0.534 & 0.478 & 0.396 & 0.513 & 0.511 & 0.518 & 0.298 & 0.439 \\
&& Casa.~\cite{yilmaz2022novo} & 0.426 & 0.341 & 0.490 & 0.478 & 0.406 & 0.521 & 0.506 & 0.537 & 0.330 & 0.448 \\
&& Ada.~\cite{xia2024adanovoadaptiveemphdenovo} & 0.467 & 0.373 & 0.593 & 0.496 & 0.431 & 0.530 & 0.546 & 0.528 & 0.372 & 0.481 \\
&& Casa.V2~\cite{yilmaz2023sequence} & 0.483 & 0.446 & 0.599 & 0.557 & 0.493 & 0.618 & 0.589 & 0.622 & 0.446 & 0.539 \\
\cline{2-13}
& \multirow{2}{*}{NAR} & Prime.~\cite{zhang2024pi} & {0.567} & 0.574 & {0.697} & {0.650} & {0.603} & {0.697} & {0.702} & {0.721} & 0.531 & {0.638} \\ 
&& \cellcolor{blue!10}\textbf{Ours} & \cellcolor{blue!10}\textbf{0.583} & \cellcolor{blue!10}\textbf{0.581} & \cellcolor{blue!10}\textbf{0.709} & \cellcolor{blue!10}\textbf{0.667} & \cellcolor{blue!10}\textbf{0.616} & \cellcolor{blue!10}\textbf{0.705} & \cellcolor{blue!10}\textbf{0.720} & \cellcolor{blue!10}\textbf{0.736} & \cellcolor{blue!10}\textbf{0.549} & \cellcolor{blue!10}\textbf{0.653} \\ 
\bottomrule 
\end{tabular}
}
\caption{Comparison of the performance on the 9-species-V1 benchmark datasets. The models are categorized by their architecture type: DB represents Database, AR stands for Autoregressive Generation, and NAR denotes Non-Autoregressive Generation. The bold font indicates the best performance.}\label{tab:testV1}
\end{threeparttable}
\end{table*}

\section{Experiments}

\subsection{Datasets}
To conduct our study, we utilized three distinct datasets in alignment with previous research for fair comparisons: MassIVE-KB~\cite{wang2018assembling}, 9-species-V1~\cite{tran2017novo}, and 9-species-V2~\cite{yilmaz2023sequence}.
MassIVE-KB is a comprehensive collection of human proteomic data, including a high-quality subset of more than 30 million peptide spectrum matches (PSMs). We then evaluated our model on the 9-species benchmark test set, which has been used in all previous de novo sequencing work and serves as a comprehensive evaluation dataset with diverse spectrum-peptide data distributions across nine species. The revised version, 9-species-V2, includes more data samples for each species and enforces a stricter annotation process for higher data quality.



\noindent\textbf{Evaluation Metrics.}
To evaluate our model's prediction accuracy, we used metrics at both amino acid and peptide levels as established by previous research. At the amino acid level, we calculated the count of correctly predicted amino acids, $N_{\text{match}}^a$, using criteria based on mass deviations~\cite{yilmaz2022novo}. Amino acid Precision was determined by $N_{\text{match}}^a / N_{\text{pred}}^a$. At the peptide level, we considered a peptide accurately predicted if all its constituent amino acids matched their ground truth counterparts. We denote $N_{\text{match}}^{\text{pep}}$ as the number of peptides with all amino acids correctly matched in a given dataset. Peptide recall was then defined as $N_{\text{match}}^{\text{pep}} / N_{\text{all}}^{\text{pep}}$, where $N_{\text{all}}^{\text{pep}}$ represents the total number of peptides in the dataset. These metrics are essential for quantifying the performance of our predictive algorithms in mass spectrometry data analysis.


\begin{figure*}[h]
\centering
\includegraphics[width=1\linewidth, height=0.2\textheight]{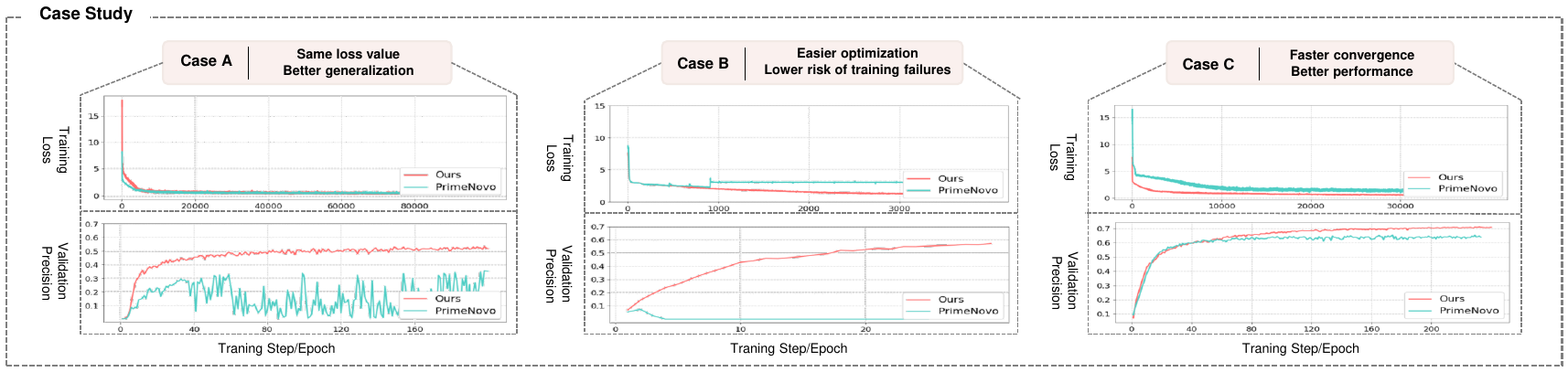}

\caption{A case study showing three types of training failures that frequently happened during the training of NAT peptide sequencing models. RefineNovo can successfully mitigate training problems in traditional NAT models. Note that the plots show training on different datasets rather than one single training for RefineNovo, and validation does not involve looking at true tokens. The training data logging was done and supported by Neptune.AI. }
\label{fig:case}
\end{figure*}

\noindent\textbf{Baselines.}
To rigorously assess our model's performance, we conducted a comparative analysis against several state-of-the-art methods. The baselines are categorized by architecture: Peaks~\cite{ma2003peaks}, representing the Database (DB) approach. The Autoregressive (AR) methods include DeepNovo~\cite{tran2017novo}, which integrates CNN and LSTM architectures; PointNovo~\cite{qiao2021computationally}, which processes mass spectrometry data across varying resolutions without increased computational complexity; Casanovo~\cite{yilmaz2022novo}, a transformer-based model; CasanovoV2~\cite{yilmaz2023sequence}, which enhances sequencing accuracy by incorporating beam search; and AdaNovo~\cite{xia2024adanovoadaptiveemphdenovo}, which leverages Conditional Mutual Information to achieve better AR performance. Lastly, PrimeNovo~\cite{zhang2024pi}, the first Non-Autoregressive Generation (NAR) method, sets a new benchmark for peptide sequencing precision.

\noindent\textbf{Model Details.} We starts by transforming all inputs, such as peaks, precursors, peptides, and amino acids, into a 400-dimensional embedding space, which serves as the foundation for further processing within the model.
Built upon this embedding space, the model's architecture features a 9-layer Transformer, where each layer contains eight attention heads. The feedforward network across all attention layers has a dimension of 1024.
To optimize the model's performance, spectra were processed with a batch size of 1600 during the training phase. An initial learning rate of 4e-4 was implemented, which was gradually increased to the target peak within the first epoch and then followed a cosine decay schedule to ensure a controlled reduction over time. Model parameters were optimized using the AdamW optimizer~\cite{kingma2014adam}.
This strategic learning rate adjustment was crucial during the 30 epochs of training, which were conducted using eight A100 GPUs. The model's hyperparameters, such as the number of layers, embedding dimensions, attention heads count, and learning rate scheduling strategy, were consistently applied as default settings for all subsequent downstream experiments unless modifications were necessary. More information on its implementation details can be referred to the code in our provided link. 

\subsection{Results}

\noindent \textbf{Performance on 9-species-V1 Benchmark Dataset.}
In this study, we assess the performance of various models on the 9-species-V1 benchmark dataset. The experimental results, summarized in Table \ref{tab:testV1}, indicate that our proposed model demonstrates superior performance in amino acid precision across most tested species, achieving an average precision of 0.800. Furthermore, our model excels in peptide recall, with an average recall rate of 0.653.

Notably, our model surpasses all other methods in both amino acid precision and peptide recall for 8 out of the 9 species tested. In these species, our model achieves the highest amino acid precision, ranging from 0.780 to 0.854, and the highest peptide recall, ranging from 0.549 to 0.736. This performance underscores the robustness and generalizability of our approach across a diverse set of species. For the human species, our NAR model achieves an amino acid precision of 0.730 and a peptide recall of 0.581. Although these results are slightly lower than those of the state-of-the-art AR model ContraNovo, they significantly narrow the gap between NAR and AR models.

Overall, the superior performance of our model demonstrates its capability to effectively identify peptides in complex mass spectrometry data, thereby highlighting the potential of our approach to advance proteomics research.

\noindent \textbf{Performance on 9-species-V2 Benchmark Dataset.}
We comprehensively evaluate our NAT model's performance against state-of-the-art baselines on the 9-species-V2 dataset, as detailed in Table \ref{tab:testV2}. Our model consistently excels in amino acid precision and peptide recall across most species. Specifically, it achieves the highest amino acid precision for all 9 species, averaging 0.907, surpassing previous NAT and AT models. In peptide recall, our model leads in 8 species, with an average of 0.790, also outperforming NAT baselines. Although ContraNovo marginally outperforms us in human recall, our model demonstrates its generalizability and overall superiority. These results confirm our model's robustness in peptide identification.

\begin{table*}[!htbp]
\setlength{\belowcaptionskip}{2mm}
\centering
\begin{threeparttable}
\setlength{\tabcolsep}{0.8mm} 
\renewcommand{\arraystretch}{1.2} 
\scalebox{0.8}{
\begin{tabular}{c|c|l|ccccccccc|c}
\toprule
Metrics & Architect & Methods & Mouse & Human & Yeast & M.mazei & Honeybee & Tomato & Rice bean & Bacillus & C.bacteria & \textbf{Average} \\
\midrule
\multirow{3}{*}{AA Precision}  & \multirow{1}{*}{AT} & Casa.V2~\cite{yilmaz2023sequence} & 0.813 & 0.872 & 0.915 & 0.877 & 0.823 & 0.891 & 0.891 & 0.888 & 0.791 & 0.862 \\
\cline{2-13}
& \multirow{2}{*}{NAT} & Prime.~\cite{zhang2024pi} & {0.839} & 0.893 & {0.932} & {0.908} & {0.862} & {0.909} & {0.931} & {0.921} & {0.827} & 0.891 \\
 & & \cellcolor{blue!10}\textbf{Ours} & \cellcolor{blue!10}\textbf{0.850} & \cellcolor{blue!10}\textbf{0.921} & \cellcolor{blue!10}\textbf{0.941} & \cellcolor{blue!10}\textbf{0.921} & \cellcolor{blue!10}\textbf{0.879} & \cellcolor{blue!10}\textbf{0.916} & \cellcolor{blue!10}\textbf{0.931} & \cellcolor{blue!10}\textbf{0.942} & \cellcolor{blue!10}\textbf{0.841} & \cellcolor{blue!10}\textbf{0.907} \\
\midrule
\multirow{3}{*}{Peptide Recall}  & \multirow{1}{*}{AT} & Casa.V2~\cite{yilmaz2023sequence} & 0.555 & 0.712 & 0.837 & 0.754 & 0.669 & 0.783 & 0.772 & 0.793 & 0.558 & 0.714 \\
\cline{2-13}
& \multirow{2}{*}{NAT} &  Prime.~\cite{zhang2024pi} & {0.627} & 0.795 & {0.884} & {0.812} & {0.742} & {0.824} & {0.837} & {0.849} & {0.626} & 0.777 \\ 
 & & \cellcolor{blue!10}\textbf{Ours} & \cellcolor{blue!10}\textbf{0.637} & \cellcolor{blue!10}\textbf{0.805} & \cellcolor{blue!10}\textbf{0.895} & \cellcolor{blue!10}\textbf{0.827} & \cellcolor{blue!10}\textbf{0.762} & \cellcolor{blue!10}\textbf{0.829} & \cellcolor{blue!10}\textbf{0.862} & \cellcolor{blue!10}\textbf{0.856} & \cellcolor{blue!10}\textbf{0.637} & \cellcolor{blue!10}\textbf{0.790} \\
\bottomrule
\end{tabular}
}
\caption{Comparison of the performance on the 9-species-V2 benchmark datasets. AT stands for Autoregressive Transformer and NAT stands for non-autoregressive Transformer. }\label{tab:testV2}
\end{threeparttable}
\end{table*}

\noindent \textbf{Training Success Rate and Case Study.}
Our approach aims to ease the learning difficulty in early-stage training and improve overall learning outcomes. We conducted a comprehensive case study demonstrating how RefineNovo assists with the training process and facilitates generalization. Specifically, we showcase three types of training obstacles frequently encountered during the training of the baseline NAT model, PrimeNovo, and compare the training loss and validation accuracy with ours. We demonstrating training on two different training datasets as adopted by \citet{mao2023mitigating} and \citet{yilmaz2022novo}. All model configurations were based on the optimal settings and were trained under the same environment, batch size, and learning rate. As shown in \ref{fig:case}, PrimeNovo frequently (up to 80\% of the time on some datasets) suffers from loss explosion (Case B), heavily depending on the choice of random seed for good convergence. In contrast, our model eliminates the problem of loss explosion due to its easy learning objective at the beginning stage by peeking at true tokens.

When both models ensure smooth loss convergence (Case A), PrimeNovo often suffers from overfitting issues with oscillating validation accuracy, suggesting the model learns to memorize rather than generalize. RefineNovo's gradual training strategy leads to much smoother validation accuracy.
Moreover, RefineNovo often exhibits higher learning speed and generalization outcomes (Case C), demonstrating the superiority of the proposed approach in training.

Lastly, we systematically evaluate training stability by randomly selecting 20 different subsets from the MassiveKB dataset and training both PrimeNovo and our model on each split. Among the 20 training runs, PrimeNovo fails in 18 cases due to loss explosion or extreme overfitting, while our model fails only once. This demonstrates that our approach is 90\% more likely to result in successful training, establishing a significantly more stable and reliable training paradigm.



\begin{table}[tb]  
\setlength{\belowcaptionskip}{2mm}  
\centering  
\begin{threeparttable}  
\setlength{\tabcolsep}{4mm} 
\renewcommand{\arraystretch}{1.2} 
\scalebox{0.9}
{
\begin{tabular}{c|cccc}  
\toprule  
\multirow{2}{*}{Methods} & \multicolumn{4}{c}{Amino Acid Precision} \\  
\cmidrule(lr){2-5}  
 & M(O) & Q & F & K \\  
\midrule  
Casa.V2 & 0.463 & 0.648 & 0.678 & 0.689 \\ 
Prime. & 0.578 & 0.770 & 0.806 & 0.800 \\  
\cellcolor{blue!10}\textbf{Ours} & \cellcolor{blue!10}\textbf{0.600} & \cellcolor{blue!10}\textbf{0.782} & \cellcolor{blue!10}\textbf{0.816} & \cellcolor{blue!10}\textbf{0.810} \\  
\bottomrule  
\end{tabular}  
}
\caption{Comparison of precision for amino acids with similar masses.}  
\label{tab:similar}  
\end{threeparttable}  

\end{table}

\noindent \textbf{Performance of RefineNovo for Similar Mass Amino Acids.}
Distinguishing amino acids with similar molecular weights is crucial for accurate peptide sequencing. Glutamine (Q) and Lysine (K) differ by just 0.036385 Da, while oxidized Methionine (Met(O)) and Phenylalanine (F) have nearly identical weights, challenging algorithmic differentiation. We rigorously evaluated {\modelname}'s proficiency in predicting these amino acids, aiming to ascertain its effectiveness in intricate cases. Figure \ref{tab:similar} highlights {\modelname}'s outstanding performance, consistently achieving superior accuracy in identifying amino acids with minimal mass differences, demonstrating its unparalleled precision.

\noindent \textbf{Ablation Study.}
Table~\ref{tab:ablation} presents an ablation study assessing the impact of three key components on our model's performance. Our baseline is naive NAT model, PrimeNovo, using simple ctc loss. Introducing the curriculum learning with a fixed mask ratio of 0.7 results in a performance decline, indicating that a fixed mask ratio may not be optimal. Adding iterative refinement with a fixed mask ratio yields a slight improvement, suggesting that iterative refinement can help correct errors. When combining the curriculum learning with difficulty annealing (dynamic masking ratio), we observe a significant enhancement over baseline model. This proves the importance of annealing ratio in adjusting learning curves. When all three components are integrated, our model achieves the highest performance. These results underscore the importance of each component and their synergistic effect, allowing our model to generate highly accurate peptide sequences.

\begin{table}[tb]
\setlength{\belowcaptionskip}{2mm}
\centering
\begin{threeparttable}
\setlength{\tabcolsep}{1.2mm} 
\renewcommand{\arraystretch}{1} 
\scalebox{0.9}
{
\begin{tabular}{ccc|c|c}
\toprule
 Curriculum & Iterative & Difficulty  & Amino Acid & Peptide\\
 Learning & Refinement & Annealing  & Precision & Recall\\
\midrule
 &  &    & 0.788 & 0.638\\
  \Checkmark &  &  & 0.733  & 0.558\\
 \Checkmark & \Checkmark &  & 0.742  & 0.571\\
\Checkmark &  & \Checkmark   & 0.793 & 0.645\\
 \cellcolor{blue!10}\Checkmark & \cellcolor{blue!10}\Checkmark & \cellcolor{blue!10}\Checkmark  & \cellcolor{blue!10}\textbf{0.800} & \cellcolor{blue!10}\textbf{0.653} \\
\bottomrule
\end{tabular}
}
\caption{Results of the ablation study showing the effects of three key components on {\modelname}'s final performance.}
\label{tab:ablation}
\end{threeparttable}

\end{table}

\section{Conclusion}

In conclusion, we introduce {RefineNovo}, a novel and effective approach to de novo peptide sequencing that sets a new performance benchmark in the field. Our model achieves superior performance and consistently surpasses previous deep learning models across all evaluated metrics. The model's enhanced training stability and the innovative integration of iterative refinement with the learnt curriculum embedding  are key contributors to its effectiveness. These advancements establish {\modelname} as a valuable tool for proteomics research and other downstream tasks. 

\section*{Impact Statement}
Peptide sequencing plays a fundamental role in proteomics, yet existing computational methods face significant challenges in efficiency, accuracy, and robustness. Our work introduces a novel curriculum learning framework tailored for non-autoregressive Transformers (NATs), addressing the longstanding difficulties of training stability and convergence in CTC-based models. By dynamically adjusting learning difficulty and integrating an iterative refinement strategy, our approach not only enhances model generalization but also significantly reduces training failures, improving both sequence prediction accuracy and reliability.

This research bridges the gap between structured learning paradigms and modern deep learning architectures, paving the way for more efficient peptide sequencing pipelines. Our methodology has the potential to accelerate discoveries in proteomics, drug development, and biomolecular analysis, where high-throughput, accurate sequencing is crucial. By releasing our implementation and trained models, we aim to provide a scalable, reproducible, and widely applicable solution to the broader scientific community.

\section*{Acknowledgement}
This project was partially supported by Shanghai Artificial Intelligence Laboratory (S.S.). This work is partially supported by Netmind.AI and ProtagoLabs Inc. This work is also partially supported by CURE (Hui-Chun Chin and Tsung-Dao Lee Chinese Undergraduate Research Endowment) (24924), and National Undergraduate Training Program on Innovation and Entrepteneurship grant(24924).
\bibliography{example_paper}

\begin{thebibliography}{64}
\providecommand{\natexlab}[1]{#1}
\providecommand{\url}[1]{\texttt{#1}}
\expandafter\ifx\csname urlstyle\endcsname\relax
  \providecommand{\doi}[1]{doi: #1}\else
  \providecommand{\doi}{doi: \begingroup \urlstyle{rm}\Url}\fi

\bibitem[Abramson et~al.(2024)Abramson, Adler, Dunger, Evans, Green, Pritzel, Ronneberger, Willmore, Ballard, Bambrick, et~al.]{abramson2024accurate}
Abramson, J., Adler, J., Dunger, J., Evans, R., Green, T., Pritzel, A., Ronneberger, O., Willmore, L., Ballard, A.~J., Bambrick, J., et~al.
\newblock Accurate structure prediction of biomolecular interactions with alphafold 3.
\newblock \emph{Nature}, pp.\  1--3, 2024.

\bibitem[Aebersold \& Mann(2003)Aebersold and Mann]{aebersold2003mass}
Aebersold, R. and Mann, M.
\newblock Mass spectrometry-based proteomics.
\newblock \emph{Nature}, 422\penalty0 (6928):\penalty0 198--207, 2003.

\bibitem[Bao et~al.(2021)Bao, Huang, Xiao, Wang, Dai, and Chen]{bao2021non}
Bao, Y., Huang, S., Xiao, T., Wang, D., Dai, X., and Chen, J.
\newblock Non-autoregressive translation by learning target categorical codes.
\newblock \emph{arXiv preprint arXiv:2103.11405}, 2021.

\bibitem[Chen et~al.(2020)Chen, Hou, Tanner, and Cheng]{chen2020bioinformatics}
Chen, C., Hou, J., Tanner, J.~J., and Cheng, J.
\newblock Bioinformatics methods for mass spectrometry-based proteomics data analysis.
\newblock \emph{International journal of molecular sciences}, 21\penalty0 (8):\penalty0 2873, 2020.

\bibitem[Ding et~al.(2020)Ding, Wang, Wu, Tao, and Tu]{ding2020context}
Ding, L., Wang, L., Wu, D., Tao, D., and Tu, Z.
\newblock Context-aware cross-attention for non-autoregressive translation.
\newblock \emph{arXiv preprint arXiv:2011.00770}, 2020.

\bibitem[Ding et~al.(2021{\natexlab{a}})Ding, Wang, Liu, Wong, Tao, and Tu]{ding2021progressive}
Ding, L., Wang, L., Liu, X., Wong, D.~F., Tao, D., and Tu, Z.
\newblock Progressive multi-granularity training for non-autoregressive translation.
\newblock \emph{arXiv preprint arXiv:2106.05546}, 2021{\natexlab{a}}.

\bibitem[Ding et~al.(2021{\natexlab{b}})Ding, Wang, Liu, Wong, Tao, and Tu]{ding2021rejuvenating}
Ding, L., Wang, L., Liu, X., Wong, D.~F., Tao, D., and Tu, Z.
\newblock Rejuvenating low-frequency words: Making the most of parallel data in non-autoregressive translation.
\newblock \emph{arXiv preprint arXiv:2106.00903}, 2021{\natexlab{b}}.

\bibitem[Eloff et~al.(2023)Eloff, Kalogeropoulos, Morell, Mabona, Jespersen, Williams, Beljouw, Skwark, Laustsen, Brouns, et~al.]{eloff2023novo}
Eloff, K., Kalogeropoulos, K., Morell, O., Mabona, A., Jespersen, J.~B., Williams, W., Beljouw, S. P.~v., Skwark, M., Laustsen, A.~H., Brouns, S.~J., et~al.
\newblock De novo peptide sequencing with instanovo: Accurate, database-free peptide identification for large scale proteomics experiments.
\newblock \emph{bioRxiv}, pp.\  2023--08, 2023.

\bibitem[Gao et~al.(2023)Gao, Chen, Wei, Jiang, and Luo]{gao2023deep}
Gao, J., Chen, J., Wei, J., Jiang, B., and Luo, A.-L.
\newblock Deep multimodal networks for m-type star classification with paired spectrum and photometric image.
\newblock \emph{Publications of the Astronomical Society of the Pacific}, 135\penalty0 (1046):\penalty0 044503, 2023.

\bibitem[Ghazvininejad et~al.(2019)Ghazvininejad, Levy, Liu, and Zettlemoyer]{ghazvininejad2019mask}
Ghazvininejad, M., Levy, O., Liu, Y., and Zettlemoyer, L.
\newblock Mask-predict: Parallel decoding of conditional masked language models.
\newblock \emph{arXiv preprint arXiv:1904.09324}, 2019.

\bibitem[Ghazvininejad et~al.(2020)Ghazvininejad, Karpukhin, Zettlemoyer, and Levy]{ghazvininejad2020aligned}
Ghazvininejad, M., Karpukhin, V., Zettlemoyer, L., and Levy, O.
\newblock Aligned cross entropy for non-autoregressive machine translation.
\newblock In \emph{International Conference on Machine Learning}, pp.\  3515--3523. PMLR, 2020.

\bibitem[Graves \& Jaitly(2014)Graves and Jaitly]{graves2014towards}
Graves, A. and Jaitly, N.
\newblock Towards end-to-end speech recognition with recurrent neural networks.
\newblock In \emph{International conference on machine learning}, pp.\  1764--1772. PMLR, 2014.

\bibitem[Graves et~al.(2006)Graves, Fern{\'a}ndez, Gomez, and Schmidhuber]{graves2006connectionist}
Graves, A., Fern{\'a}ndez, S., Gomez, F., and Schmidhuber, J.
\newblock Connectionist temporal classification: labelling unsegmented sequence data with recurrent neural networks.
\newblock In \emph{Proceedings of the 23rd international conference on Machine learning}, pp.\  369--376, 2006.

\bibitem[Gu et~al.(2017)Gu, Bradbury, Xiong, Li, and Socher]{gu2017non}
Gu, J., Bradbury, J., Xiong, C., Li, V.~O., and Socher, R.
\newblock Non-autoregressive neural machine translation.
\newblock \emph{arXiv preprint arXiv:1711.02281}, 2017.

\bibitem[Guo et~al.(2019)Guo, Tan, He, Qin, Xu, and Liu]{guo2019non}
Guo, J., Tan, X., He, D., Qin, T., Xu, L., and Liu, T.-Y.
\newblock Non-autoregressive neural machine translation with enhanced decoder input.
\newblock In \emph{Proceedings of the AAAI conference on artificial intelligence}, volume~33, pp.\  3723--3730, 2019.

\bibitem[Guo et~al.(2020)Guo, Xu, and Chen]{guo2020jointly}
Guo, J., Xu, L., and Chen, E.
\newblock Jointly masked sequence-to-sequence model for non-autoregressive neural machine translation.
\newblock In \emph{Proceedings of the 58th Annual Meeting of the Association for Computational Linguistics}, pp.\  376--385, 2020.

\bibitem[Hayes et~al.(2024)Hayes, Rao, Akin, Sofroniew, Oktay, Lin, Verkuil, Tran, Deaton, Wiggert, et~al.]{hayes2024simulating}
Hayes, T., Rao, R., Akin, H., Sofroniew, N.~J., Oktay, D., Lin, Z., Verkuil, R., Tran, V.~Q., Deaton, J., Wiggert, M., et~al.
\newblock Simulating 500 million years of evolution with a language model.
\newblock \emph{bioRxiv}, pp.\  2024--07, 2024.

\bibitem[Huang et~al.(2022{\natexlab{a}})Huang, Zhou, Za{\"\i}ane, Mou, and Li]{huang2022non}
Huang, C., Zhou, H., Za{\"\i}ane, O.~R., Mou, L., and Li, L.
\newblock Non-autoregressive translation with layer-wise prediction and deep supervision.
\newblock In \emph{Proceedings of the AAAI conference on artificial intelligence}, pp.\  10776--10784, 2022{\natexlab{a}}.

\bibitem[Huang et~al.(2022{\natexlab{b}})Huang, Zhou, Liu, Li, and Huang]{huang2022directed}
Huang, F., Zhou, H., Liu, Y., Li, H., and Huang, M.
\newblock Directed acyclic transformer for non-autoregressive machine translation.
\newblock In \emph{International Conference on Machine Learning}, pp.\  9410--9428. PMLR, 2022{\natexlab{b}}.

\bibitem[Huang et~al.(2022{\natexlab{c}})Huang, Perez, and Volkovs]{huang2022improving}
Huang, X.~S., Perez, F., and Volkovs, M.
\newblock Improving non-autoregressive translation models without distillation.
\newblock In \emph{International Conference on Learning Representations}, 2022{\natexlab{c}}.

\bibitem[Jin et~al.(2023)Jin, Xu, Zhang, Ling, Dong, Ouyang, Gao, Chang, and Sun]{jin2023contranovo}
Jin, Z., Xu, S., Zhang, X., Ling, T., Dong, N., Ouyang, W., Gao, Z., Chang, C., and Sun, S.
\newblock Contranovo: a contrastive learning approach to enhance de novo peptide sequencing. arxiv.
\newblock \emph{arXiv preprint arXiv:2312.11584}, 2023.

\bibitem[Jin et~al.(2024)Jin, Xu, Zhang, Ling, Dong, Ouyang, Gao, Chang, and Sun]{jin2024contranovo}
Jin, Z., Xu, S., Zhang, X., Ling, T., Dong, N., Ouyang, W., Gao, Z., Chang, C., and Sun, S.
\newblock Contranovo: A contrastive learning approach to enhance de novo peptide sequencing.
\newblock In \emph{Proceedings of the AAAI Conference on Artificial Intelligence}, pp.\  144--152, 2024.

\bibitem[Jumper et~al.(2021)Jumper, Evans, Pritzel, Green, Figurnov, Ronneberger, Tunyasuvunakool, Bates, {\v{Z}}{\'\i}dek, Potapenko, et~al.]{jumper2021highly}
Jumper, J., Evans, R., Pritzel, A., Green, T., Figurnov, M., Ronneberger, O., Tunyasuvunakool, K., Bates, R., {\v{Z}}{\'\i}dek, A., Potapenko, A., et~al.
\newblock Highly accurate protein structure prediction with alphafold.
\newblock \emph{Nature}, 596\penalty0 (7873):\penalty0 583--589, 2021.

\bibitem[Karunratanakul et~al.(2019)Karunratanakul, Tang, Speicher, Chuangsuwanich, and Sriswasdi]{Karunratanakul2019}
Karunratanakul, K., Tang, H.-Y., Speicher, D.~W., Chuangsuwanich, E., and Sriswasdi, S.
\newblock {Uncovering Thousands of New Peptides with Sequence-Mask-Search Hybrid De Novo Peptide Sequencing Framework.}
\newblock \emph{Molecular \& cellular proteomics : MCP}, 18\penalty0 (12):\penalty0 2478--2491, dec 2019.
\newblock ISSN 1535-9484 (Electronic).

\bibitem[Kingma \& Ba(2014)Kingma and Ba]{kingma2014adam}
Kingma, D.~P. and Ba, J.
\newblock Adam: A method for stochastic optimization.
\newblock \emph{arXiv preprint arXiv:1412.6980}, 2014.

\bibitem[LeCun et~al.(2015)LeCun, Bengio, and Hinton]{lecun2015deep}
LeCun, Y., Bengio, Y., and Hinton, G.
\newblock Deep learning.
\newblock \emph{nature}, 521\penalty0 (7553):\penalty0 436--444, 2015.

\bibitem[Lin et~al.(2023)Lin, Akin, Rao, Hie, Zhu, Lu, Smetanin, Verkuil, Kabeli, Shmueli, et~al.]{lin2023evolutionary}
Lin, Z., Akin, H., Rao, R., Hie, B., Zhu, Z., Lu, W., Smetanin, N., Verkuil, R., Kabeli, O., Shmueli, Y., et~al.
\newblock Evolutionary-scale prediction of atomic-level protein structure with a language model.
\newblock \emph{Science}, 379\penalty0 (6637):\penalty0 1123--1130, 2023.

\bibitem[Liu et~al.(2023)Liu, Ye, Li, and Tang]{liu2023accurate}
Liu, K., Ye, Y., Li, S., and Tang, H.
\newblock Accurate de novo peptide sequencing using fully convolutional neural networks.
\newblock \emph{Nature Communications}, 14\penalty0 (1):\penalty0 7974, 2023.

\bibitem[Liu et~al.(2022)Liu, Zhang, and Mou]{liu2022character}
Liu, P., Zhang, X., and Mou, L.
\newblock A character-level length-control algorithm for non-autoregressive sentence summarization.
\newblock \emph{Advances in Neural Information Processing Systems}, 35:\penalty0 29101--29112, 2022.

\bibitem[Ma et~al.(2003)Ma, Zhang, Hendrie, Liang, Li, Doherty-Kirby, and Lajoie]{ma2003peaks}
Ma, B., Zhang, K., Hendrie, C., Liang, C., Li, M., Doherty-Kirby, A., and Lajoie, G.
\newblock Peaks: powerful software for peptide de novo sequencing by tandem mass spectrometry.
\newblock \emph{Rapid communications in mass spectrometry}, 17\penalty0 (20):\penalty0 2337--2342, 2003.

\bibitem[Ma et~al.(2019)Ma, Zhou, Li, Neubig, and Hovy]{ma2019flowseq}
Ma, X., Zhou, C., Li, X., Neubig, G., and Hovy, E.
\newblock Flowseq: Non-autoregressive conditional sequence generation with generative flow.
\newblock \emph{arXiv preprint arXiv:1909.02480}, 2019.

\bibitem[Mao et~al.(2023)Mao, Zhang, Xin, and Li]{mao2023mitigating}
Mao, Z., Zhang, R., Xin, L., and Li, M.
\newblock Mitigating the missing-fragmentation problem in de novo peptide sequencing with a two-stage graph-based deep learning model.
\newblock \emph{Nature Machine Intelligence}, 5\penalty0 (11):\penalty0 1250--1260, 2023.

\bibitem[Muth et~al.(2018)Muth, Hartkopf, Vaudel, and Renard]{muth2018potential}
Muth, T., Hartkopf, F., Vaudel, M., and Renard, B.~Y.
\newblock A potential golden age to come—current tools, recent use cases, and future avenues for de novo sequencing in proteomics.
\newblock \emph{Proteomics}, 18\penalty0 (18):\penalty0 1700150, 2018.

\bibitem[Ng et~al.(2023)Ng, Zhou, and Yao]{ng2023algorithms}
Ng, C. C.~A., Zhou, Y., and Yao, Z.-P.
\newblock Algorithms for de-novo sequencing of peptides by tandem mass spectrometry: A review.
\newblock \emph{Analytica Chimica Acta}, pp.\  341330, 2023.

\bibitem[Qian et~al.(2020)Qian, Zhou, Bao, Wang, Qiu, Zhang, Yu, and Li]{qian2020glancing}
Qian, L., Zhou, H., Bao, Y., Wang, M., Qiu, L., Zhang, W., Yu, Y., and Li, L.
\newblock Glancing transformer for non-autoregressive neural machine translation.
\newblock \emph{arXiv preprint arXiv:2008.07905}, 2020.

\bibitem[Qiao et~al.(2021)Qiao, Tran, Xin, Chen, Li, Shan, and Ghodsi]{qiao2021computationally}
Qiao, R., Tran, N.~H., Xin, L., Chen, X., Li, M., Shan, B., and Ghodsi, A.
\newblock Computationally instrument-resolution-independent de novo peptide sequencing for high-resolution devices.
\newblock \emph{Nature Machine Intelligence}, 3\penalty0 (5):\penalty0 420--425, 2021.

\bibitem[Qiu et~al.(2025)Qiu, Wei, Zhang, Xu, Zou, Jin, Gao, Dong, and Sun]{qiu2025universal}
Qiu, Z., Wei, J., Zhang, X., Xu, S., Zou, K., Jin, Z., Gao, Z., Dong, N., and Sun, S.
\newblock Universal biological sequence reranking for improved de novo peptide sequencing.
\newblock \emph{arXiv preprint arXiv:2505.17552}, 2025.

\bibitem[Saharia et~al.(2020)Saharia, Chan, Saxena, and Norouzi]{saharia2020non}
Saharia, C., Chan, W., Saxena, S., and Norouzi, M.
\newblock Non-autoregressive machine translation with latent alignments.
\newblock \emph{arXiv preprint arXiv:2004.07437}, 2020.

\bibitem[Sahoo et~al.(2024)Sahoo, Arriola, Schiff, Gokaslan, Marroquin, Chiu, Rush, and Kuleshov]{sahoo2024simple}
Sahoo, S., Arriola, M., Schiff, Y., Gokaslan, A., Marroquin, E., Chiu, J., Rush, A., and Kuleshov, V.
\newblock Simple and effective masked diffusion language models.
\newblock \emph{Advances in Neural Information Processing Systems}, 37:\penalty0 130136--130184, 2024.

\bibitem[Savinov et~al.(2021)Savinov, Chung, Binkowski, Elsen, and Oord]{savinov2021step}
Savinov, N., Chung, J., Binkowski, M., Elsen, E., and Oord, A. v.~d.
\newblock Step-unrolled denoising autoencoders for text generation.
\newblock \emph{arXiv preprint arXiv:2112.06749}, 2021.

\bibitem[Shao et~al.(2020)Shao, Zhang, Feng, Meng, and Zhou]{shao2020minimizing}
Shao, C., Zhang, J., Feng, Y., Meng, F., and Zhou, J.
\newblock Minimizing the bag-of-ngrams difference for non-autoregressive neural machine translation.
\newblock In \emph{Proceedings of the AAAI conference on artificial intelligence}, volume~34, pp.\  198--205, 2020.

\bibitem[Shao et~al.(2022)Shao, Wu, and Feng]{shao2022one}
Shao, C., Wu, X., and Feng, Y.
\newblock One reference is not enough: Diverse distillation with reference selection for non-autoregressive translation.
\newblock \emph{arXiv preprint arXiv:2205.14333}, 2022.

\bibitem[Shu et~al.(2020)Shu, Lee, Nakayama, and Cho]{shu2020latent}
Shu, R., Lee, J., Nakayama, H., and Cho, K.
\newblock Latent-variable non-autoregressive neural machine translation with deterministic inference using a delta posterior.
\newblock In \emph{Proceedings of the aaai conference on artificial intelligence}, volume~34, pp.\  8846--8853, 2020.

\bibitem[Song et~al.(2021)Song, Kim, and Yoon]{song2021alignart}
Song, J., Kim, S., and Yoon, S.
\newblock Alignart: Non-autoregressive neural machine translation by jointly learning to estimate alignment and translate.
\newblock \emph{arXiv preprint arXiv:2109.06481}, 2021.

\bibitem[Stern et~al.(2019)Stern, Chan, Kiros, and Uszkoreit]{stern2019insertion}
Stern, M., Chan, W., Kiros, J., and Uszkoreit, J.
\newblock Insertion transformer: Flexible sequence generation via insertion operations.
\newblock In \emph{International Conference on Machine Learning}, pp.\  5976--5985. PMLR, 2019.

\bibitem[Tran et~al.(2017)Tran, Zhang, Xin, Shan, and Li]{tran2017novo}
Tran, N.~H., Zhang, X., Xin, L., Shan, B., and Li, M.
\newblock De novo peptide sequencing by deep learning.
\newblock \emph{Proceedings of the National Academy of Sciences}, 114\penalty0 (31):\penalty0 8247--8252, 2017.

\bibitem[Wang et~al.(2018)Wang, Wang, Carver, Pullman, Cha, and Bandeira]{wang2018assembling}
Wang, M., Wang, J., Carver, J., Pullman, B.~S., Cha, S.~W., and Bandeira, N.
\newblock Assembling the community-scale discoverable human proteome.
\newblock \emph{Cell systems}, 7\penalty0 (4):\penalty0 412--421, 2018.

\bibitem[Wang et~al.(2019)Wang, Tian, He, Qin, Zhai, and Liu]{wang2019non}
Wang, Y., Tian, F., He, D., Qin, T., Zhai, C., and Liu, T.-Y.
\newblock Non-autoregressive machine translation with auxiliary regularization.
\newblock In \emph{Proceedings of the AAAI conference on artificial intelligence}, volume~33, pp.\  5377--5384, 2019.

\bibitem[Xia et~al.(2024{\natexlab{a}})Xia, Chen, Zhou, Ling, Du, Liu, and Li]{xia2024adanovo}
Xia, J., Chen, S., Zhou, J., Ling, T., Du, W., Liu, S., and Li, S.~Z.
\newblock Adanovo: Adaptive$\backslash$emph $\{$De Novo$\}$ peptide sequencing with conditional mutual information.
\newblock \emph{arXiv preprint arXiv:2403.07013}, 2024{\natexlab{a}}.

\bibitem[Xia et~al.(2024{\natexlab{b}})Xia, Chen, Zhou, Ling, Du, Liu, and Li]{xia2024adanovoadaptiveemphdenovo}
Xia, J., Chen, S., Zhou, J., Ling, T., Du, W., Liu, S., and Li, S.~Z.
\newblock Adanovo: Adaptive \emph{De Novo} peptide sequencing with conditional mutual information, 2024{\natexlab{b}}.
\newblock URL \url{https://arxiv.org/abs/2403.07013}.

\bibitem[Xia et~al.(2024{\natexlab{c}})Xia, Liu, Zhou, Chen, Xiang, Liu, Liu, and Li]{xia2024bridging}
Xia, J., Liu, S., Zhou, J., Chen, S., Xiang, H., Liu, Z., Liu, Y., and Li, S.~Z.
\newblock Bridging the gap between database search and de novo peptide sequencing with searchnovo.
\newblock \emph{bioRxiv}, pp.\  2024--10, 2024{\natexlab{c}}.

\bibitem[Xiao et~al.(2023)Xiao, Wu, Guo, Li, Zhang, Qin, and Liu]{xiao2023survey}
Xiao, Y., Wu, L., Guo, J., Li, J., Zhang, M., Qin, T., and Liu, T.-y.
\newblock A survey on non-autoregressive generation for neural machine translation and beyond.
\newblock \emph{IEEE Transactions on Pattern Analysis and Machine Intelligence}, 45\penalty0 (10):\penalty0 11407--11427, 2023.

\bibitem[Yang et~al.(2019)Yang, Chi, Zeng, Zhou, and He]{yang2019pnovo}
Yang, H., Chi, H., Zeng, W.-F., Zhou, W.-J., and He, S.-M.
\newblock pnovo 3: precise de novo peptide sequencing using a learning-to-rank framework.
\newblock \emph{Bioinformatics}, 35\penalty0 (14):\penalty0 i183--i190, 2019.

\bibitem[Yang et~al.(2024)Yang, Ling, Sun, Liang, Xu, Huang, Xie, He, Li, He, et~al.]{yang2024introducing}
Yang, T., Ling, T., Sun, B., Liang, Z., Xu, F., Huang, X., Xie, L., He, Y., Li, L., He, F., et~al.
\newblock Introducing $\pi$-helixnovo for practical large-scale de novo peptide sequencing.
\newblock \emph{Briefings in Bioinformatics}, 25\penalty0 (2):\penalty0 bbae021, 2024.

\bibitem[Yilmaz et~al.(2022)Yilmaz, Fondrie, Bittremieux, Oh, and Noble]{yilmaz2022novo}
Yilmaz, M., Fondrie, W., Bittremieux, W., Oh, S., and Noble, W.~S.
\newblock De novo mass spectrometry peptide sequencing with a transformer model.
\newblock In \emph{International Conference on Machine Learning}, pp.\  25514--25522. PMLR, 2022.

\bibitem[Yilmaz et~al.(2024)Yilmaz, Fondrie, Bittremieux, Melendez, Nelson, Ananth, Oh, and Noble]{yilmaz2023sequence}
Yilmaz, M., Fondrie, W.~E., Bittremieux, W., Melendez, C.~F., Nelson, R., Ananth, V., Oh, S., and Noble, W.~S.
\newblock Sequence-to-sequence translation from mass spectra to peptides with a transformer model.
\newblock \emph{Nature communications}, 15\penalty0 (1):\penalty0 6427, 2024.

\bibitem[Yin et~al.(2023)Yin, Kaddour, Zhang, Nie, Liu, Kong, and Liu]{yin2023ttida}
Yin, Y., Kaddour, J., Zhang, X., Nie, Y., Liu, Z., Kong, L., and Liu, Q.
\newblock Ttida: Controllable generative data augmentation via text-to-text and text-to-image models.
\newblock \emph{arXiv preprint arXiv:2304.08821}, 2023.

\bibitem[Zhang et~al.(2024)Zhang, Ling, Jin, Xu, Gao, Sun, Qiu, Dong, Wang, Wang, et~al.]{zhang2023pi}
Zhang, X., Ling, T., Jin, Z., Xu, S., Gao, Z., Sun, B., Qiu, Z., Dong, N., Wang, G., Wang, G., et~al.
\newblock $\pi$-primenovo: An accurate and efficient non-autoregressive deep learning model for de novo peptide sequencing.
\newblock \emph{bioRxiv}, pp.\  2024--05, 2024.

\bibitem[Zhang et~al.(2025)Zhang, Ling, Jin, Xu, Gao, Sun, Qiu, Wei, Dong, Wang, et~al.]{zhang2024pi}
Zhang, X., Ling, T., Jin, Z., Xu, S., Gao, Z., Sun, B., Qiu, Z., Wei, J., Dong, N., Wang, G., et~al.
\newblock $\pi$-primenovo: an accurate and efficient non-autoregressive deep learning model for de novo peptide sequencing.
\newblock \emph{Nature Communications}, 16\penalty0 (1):\penalty0 267, 2025.

\bibitem[Zheng et~al.(2023)Zheng, Deng, Xue, Zhou, Ye, and Gu]{zheng2023structure}
Zheng, Z., Deng, Y., Xue, D., Zhou, Y., Ye, F., and Gu, Q.
\newblock Structure-informed language models are protein designers.
\newblock In \emph{International conference on machine learning}, pp.\  42317--42338. PMLR, 2023.

\bibitem[Zhou \& Keung(2020)Zhou and Keung]{zhou2020improving}
Zhou, J. and Keung, P.
\newblock Improving non-autoregressive neural machine translation with monolingual data.
\newblock \emph{arXiv preprint arXiv:2005.00932}, 2020.

\bibitem[Zhou et~al.(2024)Zhou, Chen, Xia, Liu, Ling, Du, Liu, Yin, and Li]{zhou2024novobench}
Zhou, J., Chen, S., Xia, J., Liu, S., Ling, T., Du, W., Liu, Y., Yin, J., and Li, S.~Z.
\newblock Novobench: Benchmarking deep learning-based de novo peptide sequencing methods in proteomics.
\newblock \emph{arXiv preprint arXiv:2406.11906}, 2024.

\bibitem[Zhou et~al.(2017)Zhou, Zeng, Chi, Luo, Liu, Zhan, He, and Zhang]{zhou2017pdeep}
Zhou, X.-X., Zeng, W.-F., Chi, H., Luo, C., Liu, C., Zhan, J., He, S.-M., and Zhang, Z.
\newblock pdeep: predicting ms/ms spectra of peptides with deep learning.
\newblock \emph{Analytical chemistry}, 89\penalty0 (23):\penalty0 12690--12697, 2017.

\bibitem[Zhu et~al.(2022)Zhu, Wang, and Yan]{zhu2022non}
Zhu, M., Wang, J., and Yan, C.
\newblock Non-autoregressive neural machine translation with consistency regularization optimized variational framework.
\newblock In \emph{Proceedings of the 2022 Conference of the North American Chapter of the Association for Computational Linguistics: Human Language Technologies}, pp.\  607--617, 2022.

\end{thebibliography}
\bibliographystyle{icml2025}

\newpage
\appendix
\onecolumn

\section{Curriculum Learning}
\label{app:curriculum}
Curriculum learning improves training stability and convergence in CTC-based non-autoregressive (NAT) models by dynamically adjusting the learning difficulty based on model performance. Instead of forcing the model to learn complex sequences from the start, our method progressively increases the difficulty, ensuring a smoother optimization process. 

The approach consists of four main stages. First, we compute the best alignment path using the CTC loss function, which determines the optimal token mapping between the predicted and target sequences. Next, we extract an oracle sequence, selecting key reference tokens that serve as easier learning targets for the model. To control learning difficulty, we introduce a dynamic masking ratio, where the probability of masking increases as the model's accuracy improves. Finally, we apply conditional masking, feeding the model partially masked sequences to facilitate a structured learning process that transitions from high supervision to a more generalized sequence prediction task.

The following code provides a Python implementation of our curriculum learning approach, maintaining consistency with Algorithm~\ref{alg:ctc_curriculum}. We use PyTorch to efficiently compute CTC alignments.

\begin{lstlisting}
if mode == "train":
    with torch.no_grad():
        # Forward pass without gradient updates
        word_ins_out, tgt_tokens, _ = self._forward_step(*batch)
        nonpad_positions = tgt_tokens.ne(self.decoder.get_pad_idx())
        target_lens = nonpad_positions.sum(1)
        pred_tokens = word_ins_out.argmax(-1)
        out_lprobs = F.log_softmax(word_ins_out, dim=-1)

        # Compute sequence lengths
        seq_lens = torch.full(
            (pred_tokens.size(0),), pred_tokens.size(1)
        ).to(self.device)

        # Compute best alignment using CTC
        best_aligns = best_alignment(
            out_lprobs.transpose(0, 1), tgt_tokens, seq_lens, 
            target_lens, self.decoder.get_blank_idx(), 
            zero_infinity=True
        )

        # Generate oracle sequence
        best_aligns_pad = torch.tensor(
            [a for a in best_aligns], device=word_ins_out.device
        )
        oracle_pos = (best_aligns_pad // 2).clip(
            max=tgt_tokens.shape[1] - 1
        )
        oracle = tgt_tokens.gather(-1, oracle_pos)
        oracle_empty = oracle.masked_fill(
            best_aligns_pad % 2 == 0, self.decoder.get_blank_idx()
        )

        # Compute dynamic masking ratio
        same_num = (pred_tokens == oracle_empty).sum(1)
        keep_prob = ((seq_lens - same_num) / seq_lens * peek_factor)
        keep_prob = keep_prob.unsqueeze(-1)

        # Generate curriculum learning mask
        keep_word_mask = (
            torch.rand(pred_tokens.shape, device=word_ins_out.device) 
            < keep_prob
        ).bool()
        glat_prev = oracle_empty.masked_fill(
            ~keep_word_mask, self.decoder.get_mask_idx()
        )
\end{lstlisting}

\section{Iterative Refinement}
\label{app:iter}
\subsection{Pseudo Code}
Algorithm~\ref{alg:iterative_refinement} outlines our iterative refinement approach. Instead of relying solely on a single-pass decoding, our method iteratively refines the generated peptide sequence by reintroducing previously decoded outputs as conditional input. At each iteration, the decoder takes the spectra and precursor information as input, along with the pseudo-label generated from the previous iteration. The model then refines its predictions by leveraging learned token embeddings, progressively improving sequence accuracy. This iterative process continues for a fixed number of steps, ensuring better alignment between the predicted and true sequences while mitigating common decoding errors.

\begin{algorithm}[h]
    \caption{Curriculum-Embedding-based Iterative Refinement}
    \label{alg:iterative_refinement}
    \begin{algorithmic}[1]
        \REQUIRE Spectra $\mathbf{spectra}$, Precursors $\mathbf{precursors}$, Decoder $\mathbf{decoder}$, Encoder $\mathbf{encoder}$, Iterations $N$
        \ENSURE Refined output sequence $\mathbf{output\_decoded}$

        \STATE Initialize $\mathbf{prev} \gets \text{None}$
        \STATE Initialize $\mathbf{output\_decoded} \gets [\ \ ]$
        
        \FOR {$i = 1$ to $N$}
            \STATE \textcolor{blue}{\# Forward pass with previously decoded sequence as input}
            \STATE $\mathbf{output\_logits}, \_, \mathbf{output\_list} \gets \mathbf{decoder}(\text{None}, \mathbf{precursors}, *\mathbf{encoder}(\mathbf{spectra}, \mathbf{precursors}), \mathbf{prev})$
            
            \STATE \textcolor{blue}{\# Decode using argmax and update previous predictions}
            \STATE $\mathbf{prev} \gets \arg\max(\mathbf{output\_logits}, -1)$
        
        \ENDFOR
        
        RETURN $\mathbf{prev}$
    \end{algorithmic}
\end{algorithm}

\subsection{Impact of Iterative Steps}
Table~\ref{tab:Iterative} illustrates the impact of iterative steps on performance metrics. The first iteration establishes a baseline with a peptide recall of 0.728 and an amino acid (AA) precision of 0.848. By the third iteration, peptide recall improves to 0.736 and AA precision to 0.854, demonstrating that our iterative optimization strategy results in more accurate outcomes. However, beyond the third iteration, both metrics plateau at 0.737 for recall and 0.855 for precision. This plateau suggests that while initial iterations yield significant improvements, further iterations provide minimal additional benefits. Therefore, we select three iterations to balance inference cost with model accuracy.

\begin{table}[h]
\setlength{\belowcaptionskip}{2mm}
\centering
\begin{threeparttable}
\setlength{\tabcolsep}{2mm} 
\renewcommand{\arraystretch}{1.1} 
\scalebox{0.9}
{
    \begin{tabular}{c|cc}
        \toprule
        Iterative Steps & AA Precision & Peptide Recall \\
        \midrule
        1 & 0.848 & 0.728  \\
        2  & 0.853 & 0.731\\
        3 & 0.854 & 0.736 \\
        4 & 0.855 & 0.737  \\
        5 & 0.855 & 0.737 \\
        10 & 0.855 & 0.737\\
        \bottomrule
    \end{tabular}
}
\caption{Effect of different beam sizes on {\modelname}.}
\label{tab:Iterative}
\end{threeparttable}
\end{table}

\section{CTC Loss Calculation for Protein Sequence}
\label{appendix:ctc_loss}

Connectionist Temporal Classification (CTC) loss is a widely used objective function in sequence-to-sequence models, particularly for tasks where the alignment between input and output sequences is unknown. In this section, we provide a detailed explanation of how CTC loss is calculated efficiently using dynamic programming~\cite{zhang2024pi,gu2017non}.

\subsection{Problem Definition}
The goal of CTC is to compute the total probability of all valid alignment paths $\mathbf{y}$ that reduce to the target sequence $\mathcal{A}$, denoted as $\Gamma(\mathbf{y}) = \mathcal{A}$. Let $\mathcal{I}$ represent the input spectrum and $\mathcal{A} = (a_1, a_2, \ldots, a_n)$ be the target sequence of amino acids. Each alignment path $\mathbf{y} = (y_1, y_2, \ldots, y_T)$ must satisfy the CTC reduction rules:
1. Consecutive identical tokens in $\mathbf{y}$ are merged.
2. Placeholder token $\epsilon$ is removed.
3. Identical tokens adjacent to $\epsilon$ are not merged.

CTC maximizes the total probability of all valid paths, given by:
\begin{equation}
P(\mathcal{A} | \mathcal{I}) = \sum_{\mathbf{y}:\Gamma(\mathbf{y}) = \mathcal{A}} P(\mathbf{y} | \mathcal{I}),
\end{equation}
where \( P(\mathbf{y} | \mathcal{I}) = \prod_{t=1}^T P(y_t | \mathcal{I}) \) under the independence assumption. However, directly enumerating all possible paths is computationally infeasible, as the number of valid paths grows exponentially with the length of $\mathcal{A}$ and $\mathbf{y}$. To address this, we use dynamic programming to efficiently compute $P(\mathcal{A} | \mathcal{I})$.

\subsection{Dynamic Programming Formulation}
We define $\alpha(\tau, r)$ as the probability of generating the first $r$ amino acids of the target sequence $\mathcal{A}$ using the first $\tau$ tokens in the alignment path $\mathbf{y}$:
\begin{equation}
\alpha(\tau, r) = P(A_{1:r} | S) = \sum_{\mathbf{y}:\Gamma(\mathbf{y}_{1:\tau}) = A_{1:r}} P(\mathbf{y} | S),
\end{equation}
where $S$ represents the input spectrum $\mathcal{I}$. This recursive relationship allows us to compute $P(\mathcal{A} | \mathcal{I})$ efficiently.

The initialization of $\alpha(\tau, r)$ is as follows:
\begin{align}
\alpha(\tau, 0) &= P(y_1 = \epsilon) \cdot P(y_2 = \epsilon) \cdots P(y_\tau = \epsilon), & \forall 1 \leq \tau \leq T, \\
\alpha(1, 1) &= P(y_1 = a_1), & \\
\alpha(1, r) &= 0, & \text{for } r > 1.
\end{align}

\subsection{Recursive Calculation}
To compute $\alpha(\tau, r)$ for $\tau > 1$ and $r \geq 1$, we decompose it based on whether the current token $y_\tau$ matches the target amino acid $a_r$, or if it contributes to a blank or repeated token. Using the law of total probability:
\begin{align}
\alpha(\tau, r) = 
\begin{cases}
\alpha(\tau-1, r) \cdot P(y_\tau = a_r), & \text{if } a_r = a_{r-1}, \\
\alpha(\tau-1, r-1) \cdot P(y_\tau = a_r), & \text{if } a_r \neq a_{r-1}, \\
\alpha(\tau-1, r) \cdot P(y_\tau = \epsilon), & \text{otherwise}.
\end{cases}
\end{align}

This recurrence efficiently aggregates the probabilities of all valid paths reducing to $A_{1:r}$.

\subsection{Loss Function}
The final CTC loss is defined as the negative log probability of the target sequence $\mathcal{A}$:
\begin{equation}
\mathcal{L}_{\text{CTC}} = -\log P(\mathcal{A} | \mathcal{I}) = -\log \alpha(T, |\mathcal{A}|).
\end{equation}

\subsection{Practical Considerations}
To ensure numerical stability during training, the log-space version of the dynamic programming equations is often used. Additionally, dynamic programming reduces the computational complexity from exponential to linear in terms of $T$ and $n$. This makes CTC loss feasible for large-scale peptide sequencing tasks.

\subsection{Illustration of CTC Path}
An example of the valid paths for a target sequence $\mathcal{A} = (A, T, C)$ and alignment length $T = 5$ is shown below:
\begin{itemize}
    \item Path 1: \( A \ A \ T \ T \ C \)
    \item Path 2: \( A \ \epsilon \ T \ \epsilon \ C \)
    \item Path 3: \( \epsilon \ A \ T \ C \ \epsilon \)
\end{itemize}
Only paths that reduce to the exact target sequence according to $\Gamma(\cdot)$ are considered valid.

Dynamic programming significantly simplifies the computation of CTC loss by avoiding explicit enumeration of paths. This approach ensures efficiency and scalability in de novo peptide sequencing tasks, enabling the model to focus on meaningful alignment paths while discarding invalid ones. The detailed recursive formulation provided here serves as the foundation for implementing robust CTC-based training for sequence generation models.

\section{Precise Mass Control (PMC) Method}
\label{appendix:pmc}

\subsection{Overview}
Precise Mass Control (PMC)~\cite{zhang2024pi} is a knapsack-like dynamic programming algorithm designed to enforce mass constraints during peptide decoding. The goal of PMC is to ensure that the total mass of the generated peptide sequence aligns with the experimentally measured precursor mass $m_{\text{pr}}$, within a predefined error tolerance $\sigma$. This ensures that the generated peptide is both accurate and physically valid, addressing challenges in non-autoregressive sequence generation where no direct mechanism exists to enforce mass constraints during decoding.

\subsection{Problem Formulation}
The PMC problem can be formalized as maximizing the total log probability of generating a peptide sequence $\mathcal{A} = (a_1, a_2, \ldots, a_n)$, subject to a mass constraint:
\begin{equation}
\max_{\mathcal{A}} \sum_{i=1}^n \log P(y_i | \mathcal{I}),
\end{equation}
where $\mathcal{I}$ represents the input spectrum, $P(y_i | \mathcal{I})$ is the model's predicted probability for amino acid $y_i$ at position $i$, and the mass constraint is given by:
\begin{equation}
m_{\text{pr}} - \sigma \leq \sum_{a_i \in \mathcal{A}} u(a_i) \leq m_{\text{pr}} + \sigma,
\end{equation}
where $u(a_i)$ is the mass of amino acid $a_i$. The goal is to find a sequence $\mathcal{A}$ that maximizes the probability while satisfying the mass constraint.

\subsection{Dynamic Programming Approach}
To solve this optimization problem, we employ a dynamic programming (DP) table $d^\ell_t$, where $t$ denotes the step index and $\ell$ denotes the mass at that step. The DP table stores the most probable sequence of amino acids that satisfies the mass constraint at each step.

\subsubsection{Initialization}
For the first decoding step ($t=1$), we initialize the DP table as follows:
\begin{equation}
d^\ell_1 =
\begin{cases}
\epsilon, & \text{if } \ell = 0, \\
y_1, & \text{if } u(y_1) \in [\ell - \sigma, \ell + \sigma], \\
\emptyset, & \text{otherwise}.
\end{cases}
\end{equation}
Here, $\epsilon$ represents the empty sequence, $y_1$ is the first amino acid in the sequence, and $u(y_1)$ denotes its mass.

\subsubsection{Recursion}
At each decoding step $t > 1$, the DP table is updated by considering three cases, depending on the nature of the new token $y_t$:
1. If $y_t = \epsilon$, the mass remains unchanged due to CTC reduction. In this case:
   \begin{equation}
   d^\ell_t = \bigcup_{\gamma \in d^{\ell}_{t-1}} \gamma \circ \epsilon,
   \end{equation}
   where $\circ$ denotes sequence concatenation.

2. If $y_t$ is identical to the previous token, the sequence remains the same, but the mass remains constant:
   \begin{equation}
   d^\ell_t = \bigcup_{\gamma \in d^{\ell}_{t-1}} \gamma \circ y_{t-1}.
   \end{equation}

3. If $y_t$ is a new token, the mass increases, and the potential sequences are updated as:
   \begin{equation}
   d^\ell_t = \bigcup_{\gamma \in d^{\ell - u(y_t)}_{t-1}} \gamma \circ y_t.
   \end{equation}

\subsubsection{Mass-Constrained Update}
To ensure the DP table does not grow excessively large, we retain only the top $B$ sequences with the highest probabilities at each step:
\begin{equation}
d^\ell_t = \text{top}_B \bigg( \bigcup_{y_t \in \mathcal{Y}} \sum_{\gamma \in d^{\ell - u(y_t)}_{t-1}} P(\gamma) \bigg),
\end{equation}
where $\mathcal{Y}$ represents the set of all amino acid tokens.

\subsection{Final Sequence Selection}
After completing all decoding steps, the final sequence is selected as the most probable sequence stored in $d^\ell_T$, where $T$ is the total decoding length:
\begin{equation}
\mathcal{A} = \arg\max_{\gamma \in d^\ell_T} P(\gamma).
\end{equation}

\section{Performance Comparison on NovoBench}
To rigorously benchmark model performance, we evaluated {\modelname} and all baseline models on NovoBench~\cite{zhou2024novobench}, a recently released de novo sequencing benchmark dataset. We adhered to the experimental setup and evaluation protocol outlined in the NovoBench publication, consistently utilizing Saccharomyces cerevisiae (\texttt{yeast}) as the test species for all comparative analyses. We directly evaluated our existing pretrained model, which was trained on MassiveKB, against the NovoBench \texttt{yeast} test sets. This approach, while not leveraging benchmark-specific training, provides a robust assessment of our model's generalization capabilities.

For a direct and equitable comparison with PrimeNovo, the prior state-of-the-art Non-Autoregressive Transformer (NAT)-based model, we utilized its publicly available weights. Both RefineNovo and PrimeNovo were evaluated under identical conditions on the NovoBench test data to ensure comparability.

We notice the relatively low performance on the 7-species dataset when directly testing with all pretrained models such as Casanovo, PrimeNovo, and RefineNovo. Upon further investigation, we found that this dataset was generated using MS equipment with precision levels significantly different from those in the MassiveKB training data. This results in a notable distribution mismatch. Nevertheless, despite this domain shift, the pretrained RefineNovo model still demonstrates clearly better performance compared to other models trained on MassiveKB. This highlights the robustness of our method under distributional variation.

\begin{table}[htbp]
\centering
\caption{Performance comparison on the NovoBench benchmark (yeast test species). Scores for models marked with * are quoted from the NovoBench paper or original publications. CV denotes cross-validation results from the original PrimeNovo paper. ``--'' indicates data not available.}
\label{tab:novobench_results}
\scalebox{0.7}{!}{%
\begin{tabular}{@{}lccc@{}}
\toprule
Model & \texttt{9Species (yeast)} & \texttt{7Species (yeast)} & \texttt{HC-PT} \\ \midrule
Casanovo * & $0.48$ & $0.12$ & $0.21$ \\
InstaNovo * & $0.53$ & --     & $0.57$ \\
AdaNovo * & $0.50$ & $0.17$ & $0.21$ \\
HelixNovo * & $0.52$ & $0.23$ & $0.21$ \\
SearchNovo * & $0.55$ & $0.26$ & $0.45$ \\
PrimeNovo-CV * & $0.58$ & --     & --     \\ \midrule
Casanovo-pretrained & $0.60$ & $0.05$ & --     \\
PrimeNovo         & $0.70$ & $0.09$ & $0.85$ \\
\textbf{RefineNovo (ours)} & $\mathbf{0.71}$ & $\mathbf{0.09}$ & $\mathbf{0.88}$ \\ \bottomrule
\end{tabular}%
}
\end{table}

\section{Relationship to Prior Work in Iterative Refinement and Difficulty Annealing} 

The development of robust generative models for sequences, particularly in complex domains like peptide generation, benefits significantly from strategies that enhance output quality and training stability. In this context, our proposed self-refining module and difficulty annealing strategy build upon established concepts while introducing specific innovations tailored to Non-Autoregressive Transformer (NAT) models and Connectionist Temporal Classification (CTC) based decoding.

\subsection{Iterative Self-Refinement} 

The principle of iteratively refining generated outputs to improve quality is a powerful paradigm. Our post-training self-refining module, integrated into the main NAT architecture, draws inspiration from multi-pass generation techniques employed in notable protein-related models like ESM-3 and AlphaFold2, which leverage iterative processing for enhanced prediction accuracy.

We acknowledge that the motivation for such refinement---improving generation quality through successive rounds of error correction and adjustment---is shared with several existing lines of research. For instance, masked language modeling approaches, including conditional masked language modeling, and discrete or masked diffusion models also employ iterative refinement~\cite{ghazvininejad2019mask,zheng2023structure, sahoo2024simple}, often by re-predicting masked or noisy portions of a sequence.

Our method, however, introduces a distinct mechanism specific to NAT models utilizing CTC. Instead of directly refining the generated token sequence, our self-refinement module operates on the \textbf{CTC path}. A CTC path represents one of many possible alignment and reduction outcomes that can produce the target label sequence. In our framework, an initial forward pass yields a CTC path. This path is then fed back into the model, allowing for iterative adjustment and refinement of this alignment representation in subsequent passes. The final, refined \textbf{CTC path} is then used for the reduction to the ultimate output sequence. This \textbf{CTC-path-centric refinement} allows the model to explore and optimize the alignment space more effectively within the NAT framework, differentiating it from methods that directly manipulate the sequence tokens during refinement. 

\subsection{Difficulty Annealing in Sequence Generation} 

Curriculum learning, or ``easy-to-hard'' training strategies, has demonstrated efficacy in various machine learning tasks, including sequence generation. Prior work has explored difficulty annealing primarily at a coarser granularity, such as at the \textbf{task level} (learning simpler tasks before more complex ones) or at the \textbf{inter-sequence level} (e.g., training on shorter or structurally simpler sequences before progressing to longer or more complex ones, as explored in some protein generation contexts~\cite{ghazvininejad2019mask}.

To the best of our knowledge, our proposed difficulty annealing strategy is among the first to implement annealing at a \textbf{within-sequence} granularity for NAT models with CTC. Our method defines and modulates difficulty \emph{within each training sequence} by controlling the amount of information exposed from its chosen CTC path during training. Each sequence effectively starts as an ``easier'' instance by revealing a greater portion of its CTC path. As training progresses, the visibility of this path information is gradually reduced. This reduction exponentially increases the learning difficulty due to the \emph{combinatorial explosion} in the number of valid CTC paths that could correspond to the same target label sequence.

This fine-grained \textbf{within-sequence} and \textbf{within-path} difficulty annealing is uniquely enabled by our CTC-sampling mechanism, which is specifically designed for this purpose. This approach plays a crucial role in stabilizing the training dynamics of our NAT model and guiding it towards more robust representations.

\end{document}